\documentclass[acmtog]{acmsmall} 

\usepackage[utf8]{inputenc} 

\usepackage[ruled]{algorithm2e}

\SetAlFnt{\small}
\SetAlCapFnt{\small}
\SetAlCapNameFnt{\small}
\SetAlCapHSkip{0pt}
\IncMargin{-\parindent}

\usepackage{booktabs}
\usepackage{array}
\usepackage{todonotes}
\usepackage[protrusion=true]{microtype}
\usepackage{afterpage}
\usepackage{pdflscape}

\usepackage{ragged2e}
\newcolumntype{P}[1]{>{\RaggedRight\arraybackslash}p{#1}}

\acmVolume{TBD}
\acmNumber{TBD}
\acmArticle{TBD}
\acmYear{2017}
\acmMonth{3}

\doi{0000001.0000001}

\issn{1234-56789}

\newcommand{\fref}[1]{Figure~\ref{#1}}
\newcommand{\sref}[1]{Section~\ref{#1}}
\newcommand{\aref}[1]{Appendix~\ref{#1}}
\newcommand{\tref}[1]{Table~\ref{#1}}

\usepackage[normalem]{ulem} 
\usepackage{xcolor}

\usepackage{ifthen}
\usepackage{amssymb}
\newboolean{showcomments}
\setboolean{showcomments}{true} 
\ifthenelse{\boolean{showcomments}}
  {\newcommand{\nb}[2]{
    \fcolorbox{gray}{yellow}{\bfseries\sffamily\scriptsize#1}
    {\sf\small$\blacktriangleright$\textit{#2}$\blacktriangleleft$}
   }
   
  }
  {\newcommand{\nb}[2]{}
   
  }

\begin{document}

\acmformat{Jan-Philipp Stegh{\"o}fer, 
H\r{a}kan Burden, 
Regina Hebig,
Gul Calikli, 
Robert Feldt, 
Imed Hammouda, 
Jennifer Horkoff, 
Eric Knauss, and 
Grischa Liebel. Involving External Stakeholders in Project Courses}

\begin{bottomstuff}
Author's addresses: Software Engineering Division, Department of Computer Science and Engineering, Chalmers $|$ University of Gothenburg, Sweden; RISE Viktoria, Sweden
\end{bottomstuff}

\author{Jan-Philipp Stegh{\"o}fer \affil{Chalmers $|$ University of Gothenburg, Sweden}
H\r{a}kan Burden \affil{RISE Viktoria, Sweden}
Regina Hebig,
Gul Calikli,
Robert Feldt \affil{Chalmers $|$ University of Gothenburg, Sweden}
Imed Hammouda \affil{Chalmers $|$ University of Gothenburg, Sweden and Mediterranean Institute of Technology, South Mediterranean University, Tunisia}
Jennifer Horkoff, Eric Knauss, and
Grischa Liebel \affil{Chalmers $|$ University of Gothenburg, Sweden}
}

\markboth{Stegh{\"o}fer et al.}{Involving External Stakeholders in Project Courses}

\title{Involving External Stakeholders in Project Courses}

\begin{abstract}
\textbf{Problem:} 
The involvement of external stakeholders in capstone projects and project courses is desirable due to its potential positive effects on the students. Capstone projects particularly profit from the inclusion of an industrial partner to make the project relevant and help students acquire professional skills. In addition, an increasing push towards education that is aligned with industry and incorporates industrial partners can be observed. However, the involvement of external stakeholders in teaching moments can create friction and could, in the worst case, lead to frustration of all involved parties.

\noindent\textbf{Contribution:}
We developed a model that allows analysing the involvement of external stakeholders in university courses both in a retrospective fashion, to gain insights from past course instances, and in a constructive fashion, to plan the involvement of external stakeholders.

\noindent\textbf{Key Concepts:}
The conceptual model and the accompanying guideline guide the teachers in their analysis of stakeholder involvement. The model is comprised of several activities (define, execute, and evaluate the collaboration). The guideline provides questions that the teachers should answer for each of these activities. In the constructive use, the model allows teachers to define an action plan based on an analysis of potential stakeholders and the pedagogical objectives. In the retrospective use, the model allows teachers to identify issues that appeared during the project and their underlying causes. Drawing from ideas of the reflective practitioner, the model contains an emphasis on reflection and interpretation of the observations made by the teacher and other groups involved in the courses.

\noindent\textbf{Key Lessons:}
Applying the model retrospectively to a total of eight courses shows that it is possible to reveal hitherto implicit risks and assumptions and to gain a better insight into the interaction between external stakeholders and students. Our empirical data reveals seven recurring risk themes that categorise the different risks appearing in the analysed courses. These themes can also be used to categorise mitigation strategies to address these risks pro-actively. Additionally, aspects not related to external stakeholders, e.g., about the interaction of the project with other courses in the study program, have been revealed. 
The constructive use of the model for one course has proved helpful in identifying action alternatives and finally deciding to not include external stakeholders in the project due to the perceived cost-benefit-ratio. 

\noindent\textbf{Implications to practice:}
Our evaluation shows that the model is viable and a useful tool that allows teachers to reason about and plan the involvement of external stakeholders in a variety of course settings, and in particular in capstone projects.
\end{abstract}

%
%


%
%



\maketitle

\section{External Stakeholders in Project Courses}

Policy makers such as the European Commission call for curriculum development \emph{``drawing on new methods of teaching and learning, so that students acquire relevant skills that enhance their employability''}~\cite{eu_quality_hie}.
To increase employability, it is necessary to understand future employers' needs. Lutz et al.~\citeyear{lutz2014undergraduate} summarize these needs under the keywords professionalism, execution competence, as well as technical knowledge and expertise.
While the latter need is classically addressed in university education by providing solid academic and theoretical foundations, professionalism and execution competence are practical skills that are based on a valuable good: experience. This, however, can only be gained by exposing students to realistic settings and problems.
Therefore, in our courses and particularly in capstone projects, we 
often complement academic education with contributions from external stakeholders, in particular stakeholders from industry.

At the same time, institutional pressure for including external stakeholders in education increases. In Sweden, e.g., traditionally one-third of the university boards' members are external stakeholders~\cite{musial2010redefining}. Musia{\l}~\citeyear{musial2010redefining} discusses the increasing involvement of external stakeholders in university boards pointing out that the involvement of non-academic stakeholders leads to a shift in the perception of higher education institutions towards a market orientation. 
Many engineering programs take even more direct input from industry, by asking external stakeholders about desired teaching outcomes. For example, Kans~\citeyear{kans2016should} conducted a survey including industry representatives, to identify aspects that should be included in a bachelors engineering program.

These factors contribute to a teacher's decision to include external stakeholders in a course, but only paint a partial picture. Indeed, there are several reported benefits for the students as well as the teachers that go beyond external forces. Including external stakeholders in educational activities is thus driven by various motivations, such as:

\begin{longitem} 
\item to allow students to interact with industrial practice~\cite{harrison1997enhancing,wohlin1999achieving}; 
\item to teach practical knowledge and soft skills~\cite{bruegge2015software};
\item to help students to acquire skills to solve real-life problems through putting the theoretical knowledge they learn in class into practice~\cite{bruegge2015software}; 
\item to enable students to learn techniques that are specific to large systems, e.g., working with and comprehending legacy code~\cite{fox2014software}, or not realistically teachable otherwise~\cite{gabrysiak2012teaching}
\item to increase students' motivation due to the chance that a project's outcome might be used in practice~\cite{williams2011five};
\item to increase credibility of education by relating to industrial practice~\cite{dagnino2014increasing};
\item to show students a diversity of opinions~\cite{dagnino2014increasing};
\item 
to put students in a good position to find high-quality jobs after graduation through networking~\cite{williams2011five,harrison1997enhancing}; and
\item to address governmental and institutional pressure for relevant knowledge and skills~\cite{eu_quality_hie}.
\end{longitem} 

Not just teachers and students have interest in the stakeholder involvement. Often, the stakeholders themselves have good reasons to be motivated, as mentioned by Harrison~\citeyear{harrison1997enhancing}. For example, they see the interaction as an option to attract future employees. Furthermore, student projects allow them to create prototypes for new business ideas, identify errors in their requirements, and benefit from student's creativity \cite{AMCIS16}. In some cases stakeholders even come with the expectation to gain a ready-for-use component for their systems.

Indeed, we observe a number of benefits in our teaching practice: student attendance and motivation can be higher when external stakeholders are involved; the students take the material more seriously when it is augmented by reports from practitioners; guest lectures and supervision by external stakeholders can decrease the workload for the teachers. These positive effects align with the reported motivations.
At the same time, including external stakeholders leads to
challenges that affect the students, the external stakeholders, and the teachers. For instance, we observe that students are frustrated when stakeholders use different terminology than teachers. We also observe that external stakeholders are frustrated when students are aiming more for academic achievements than for fulfilling stakeholder needs. And
we, as teachers, are frustrated when students do not take away the insights we want them to gain from the interaction.

Furthermore, it is difficult to thoroughly plan the involvement of external stakeholders in the courses and the various teaching moments since it is difficult to anticipate the interaction between students and external stakeholders and ensure that these interactions are constructively aligned~\cite{biggs1996enhancing} with the course objectives.

\subsection{A Model of Stakeholder Involvement}

These observations have motivated us to create a model that can be used to analyse and plan the involvement of external stakeholders in project courses. We have created this model based on our experiences in various courses across a number of programs and universities. We show how the model can be applied \emph{retrospectively}, in order to identify challenges and positive outcomes of stakeholder involvement in project courses. To a lesser extent, we also show the usefulness of the model as a planning instrument, when applying it \emph{constructively}. The main contribution of this paper is thus an empirically validated model of external stakeholder involvement that can be applied by teachers in accordance with Sch\"{o}n's reflective practitioner approach~\cite{Schon1983} to better understand the interaction with external stakeholders in their courses and the impact different assumptions, constraints, and choices have. Our reflection also produced a number of mitigation strategies that have proven successful in the past. We offer these experiences as support for teachers who find themselves in similar situations.

We have been working as a ``guild of teachers'' --- a model that has proven useful and successful in the past~\cite{steghofer2016teaching} --- to reflect on our teaching from different perspectives. This approach fosters a Scholarship of Teaching and Learning within the faculty~\cite{SoTL}. 
The goal of this reflection was to collect case studies of stakeholder interaction in project courses as empirical data,
discuss observed challenges and mitigation strategies, and
provide a model that allows to reason about the underlying mechanisms that lead to the observed outcomes. In addition, it was our goal to share our experiences in how to carry out successful collaborations, and to enable us to effectively plan future project courses with external stakeholder involvement.

\subsection{Structure of the Paper}

This paper is structured as follows: we begin by introducing relevant theoretical background and relating our own work to existing literature in \sref{sec:related-work}. Then, we detail the methodology that we have applied to arrive at our final model of stakeholder involvement, how we validated the model, and how we extracted mitigation strategies in \sref{sec:research-methodology}. In that section, we also detail the threats to validity and the strategies we employed to avoid them. The model itself is introduced in \sref{sec:model}, and its application to a variety of project courses is outlined in \sref{sec:application}. We detail the results of that application, the strengths and weaknesses of the model, as well as the lessons learned in \sref{sec:results}. \sref{sec:constructive} gives a brief insight into our (limited) experience of using the model in a constructive fashion, i.e., to plan the next iteration of our courses. We conclude the paper by discussing the contribution and our intended future work in \sref{sec:conclusion}.

\section{Related Work} 
\label{sec:related-work}






In the following, we discuss related work that addresses general systematic approaches towards course design (\sref{sec:related-work:systematic-approaches}) and integrating external stakeholders in courses (\sref{sec:related-work:integrating-stakeholders}). We then review work combining a systematic approach with the integration of stakeholders (\sref{sec:related-work:systematic-stakeholders}) before summarising the related work (\sref{sec:related-work:summary}).

\subsection{Systematic Approaches towards Course Designs}
\label{sec:related-work:systematic-approaches}
Research on higher education provides systematic approaches towards course design.
The most relevant is the ADDIE model, which helps teachers to plan and evaluate their course design in five steps (Analyse, Design, Develop, Implement, and Evaluate)~\cite{bates2016teaching,dick2006systematic,morrison2010designing}. 
The resemblance to a software development process is not accidental, as ADDIE considers the incorporation of multimedia and internet components into the course set-up. Thus, it is not surprising that the model is also criticised for the same reasons as the waterfall model --- the student perspective has a marginal role, the model does not account for the unbalanced workload of students over the term, up-front scheduling can not foresee issues during the implementation phase, and it is difficult to revise the plan in case of failure~\cite{Lembo2012}.   

Bertram~\citeyear{bertramagile}, and Rawsthorne and Lloyd~\citeyear{Rawsthorne2005} introduce the idea of agile course designs, comparing the more traditional ADDIE model to the waterfall process. In agile course designs, the course, or parts of it, are planned and created during the execution of the course in cooperation with the students~\cite{Stewart2009}. Stewart et al.\ also emphasise student team work and the collaboration between students and teachers for successful course design in this context, and detail how problem-based learning maps to agile course design. A key concept is that the teacher is a facilitator for the students to reach knowledge and understanding instead of the ``source of knowledge''. 

Both models include interesting components for course planning and refinement in general.\ However, they are not sufficient when it comes to planning and refining the interaction with external stakeholders in education.


\subsection{Integrating Stakeholders in Computer Science Teaching} 
\label{sec:related-work:integrating-stakeholders}

In the late nineties universities began with integrating real customers into project courses. Since then, a multitude of universities have followed by integrating 
start-ups~\cite{gabrysiak2012teaching}, 
NGOs and other non-profit partners~\cite{6595232,penzenstadler2014using}, 
public institutions such as libraries~\cite{Boehm1998}, or
industry partners, e.g., BMW AG~\cite{penzenstadler2013university}, Munich Airport~\cite{bruegge2008experiment}, and others~\cite{kornecki1997strengthening,daun2016project,bruegge2015software,daun2014industrial,tahmoush2009enhancing,hadfield2007crafting,rosiene2006experiences}. 
The course content varies from specific topics, such as requirements engineering and software project management, to software engineering in general.\ 

Some of these publications even discuss multiple course iterations and attempt to measure benefits of involving external stakeholders. 
For example, Daun et al.\ investigate the benefits for students' motivation~\cite{daun2014industrial} and Bruegge et al.\ investigate how students' skills in modelling and programming improve~\cite{bruegge2015software}.

Most of the reports agree on the value of the interaction between students and external stakeholders. For example, 
Tahmoush et al.\ conclude that the confrontation with industrial code bases is beneficial for the students to learn about maintenance and upgrading old code~\cite{tahmoush2009enhancing}. Dagnino~\citeyear{dagnino2014increasing} reports on a joined effort of ABB and the North Carolina State University to create a simulated engineering environment for their students to gain skills, such as effective team work.
As a part of that they propose to include external stakeholders from industry as guest lecturers.
However, there are also hints on challenges with the integration of external stakeholders. 
For example, Kornecki et al.\ present a course where students work on an air traffic system provided by Lockheed Martin Air Traffic Management~\cite{kornecki1997strengthening}. The authors report on 
scheduling incompatibilities between academia and industry, 
lacking industrial practice of teachers (which was mitigated by giving them extra training before the project), or
difficulties for teachers to invest enough time.
Similarly, Boehm et al.\ observe mismatches between priorities of the stakeholder and students' learning experience when only parts of the students continued with the second part of the course. 
In this case the teachers decided against interrupting running students' projects to reassign project parts following the stakeholder's prioritization~\cite{Boehm1998}.

\subsection{Systematic Approaches towards Computer Science Course Designs that Integrate External Stakeholders}
\label{sec:related-work:systematic-stakeholders}

In many of the papers listed above, the authors describe their course designs in a reproducible way and some even explain how they used lessons learned to improve future course instances. 
However, there is little work that actually provides criteria or guidelines to systematically plan and reflect on stakeholder interaction. 

Wohlin and Regnell present strategies to make a masters program more relevant to industry, such as guest lectures from industry and master's theses in industry~\cite{wohlin1999achieving}. The authors conclude that there is room to improve the collaboration between industry and academia. Sedano et al.\ provide a method to classify suitability of project proposals for university courses~\cite{sedano2016green}. This includes criteria about the external stakeholders, such as the question whether responsible contact persons at stakeholder side are identified. 

Harrison goes a step further by introducing a methodology for incorporating industrial participation in project courses~\cite{harrison1997enhancing}.
This covers the identification of potential industry sponsors (i.e., the stakeholders who provide the projects), sponsor assessment criteria, and project assessment criteria. In addition, the author presents a model that describes the role and duties of the lecturer and rules for the course, e.g., uniform tasks for all teams, or periodic visits of the industry sponsor. A related approach is that by H{\"o}st et al.~\citeyear{Host2010SEThesisSupport}. Based on interviews with external stakeholders that have acted as industrial supervisors for capstone projects of software engineering master students, the authors characterise different types of projects and propose a support model structured around these project types. The interview results also highlight goals of stakeholders involved in capstone projects, barriers that the stakeholders and students see to successful projects, and concrete ways to propose capstone projects. However, the focus is on improving the projects rather than specifically on the external stakeholders.

A distinguishing factor of the contribution of Boehm et al.\ is that the authors use a process to prepare the stakeholder for the interaction with the students, i.e., by running the first cycle of the projects process without students, in order to determine the feasibility of the project and to negotiate the stakeholder involvement.

However, all of these approaches have a quite limited view of the involvement of the industrial stakeholder by regarding them only as customers.
There is no notion of 
stakeholders who act, e.g., as supervisors or mentors to the students.

\subsection{Summary}
\label{sec:related-work:summary}
To sum up, literature widely agrees that the involvement of external stakeholders in education is beneficial. However, it also becomes clear that there are challenges with that involvement. So far, only few papers address this with proposing decision support or even applying a dedicated process for trading-off stakeholders', students', and teachers' aims.
A comprehensive approach to support course planning by covering aims and external conditions is still missing.

\section{Research Methodology}
\label{sec:research-methodology}

Our methodology can be split in several parts, differentiated by their aims and by their contribution to the research questions.
We address three research questions:

\begin{description}
	\item[RQ1] How can we model the involvement of external stakeholders in course design?
	\item[RQ2] Which risk themes and mitigating strategies emerge when applying the model?
	\item[RQ3] Which are the identified strengths and weaknesses of the model? 
\end{description}

We answer \emph{RQ1} by reflecting on our own experiences of involving external stakeholders. By doing so, we develop a model of stakeholder involvement and validate it by applying it to different courses. 
To answer \emph{RQ2} and \emph{RQ3}, we apply the model to our own courses to extract risks and mitigation strategies, their assumptions, and their effects as well as benefits and shortcomings of our model. 


The development and validation of the model happened in several iterations in a form of \emph{action research} \cite{Lewin1946,Kember1992} and is detailed in \sref{sec:research-methodology:model}. 
The extraction of common risks and mitigation strategies from the derived data is detailed in \sref{sec:research-methodology:mitigation}. 
We are aware of threats to validity and employed a number of countermeasures as detailed in \sref{sec:research-methodology:threats-to-validity}.

\subsection{Action Research for the Creation and Validation of the Model}
\label{sec:research-methodology:model}

The main contribution of this paper is a model for reasoning about the involvement of external stakeholders in project courses at university level. We created this model within the ``guild of teachers'' --- a group of academics of various levels of seniority, including tenured professors and senior faculty in various stages of their career, as well as a PhD student --- in an iterative process. We discussed each iteration in the guild, and communicated changes and their reasons to all members. We then validated the updated model by applying it to a selection of courses taught by teachers in the guild. In the next iteration, we used the resulting feedback from this validation to refine the model. In total, we performed six such iterations.
Our work was guided by Brookfield's lenses~\cite{brookfield1995becoming} and Sch\"{o}n's reflective practitioner approach~\cite{Schon1983}, based on empirical data from past course instances. We also tapped into the knowledge of the guild by collecting information, discussing, and reflecting as a group. A rough overview of this process is given in \tref{tab:iterations}. We discuss our reasoning process and the model's evolution in the following.

 \begin{table}[b]
\centering
\tbl{Iterations of the model, their respective main content, and their validation.}{
\begin{tabular}{@{}P{1.4cm} P{1.6cm} P{1.6cm} P{1.8cm} P{1.6cm} P{1.6cm} P{1.7cm}@{}}
  \toprule
   & \textbf{Iteration 1} & \textbf{Iteration 2} & \textbf{Iteration 3} & \textbf{Iteration 4} & \textbf{Iteration 5} & \textbf{Iteration 6} \tabularnewline
  \midrule
\textbf{Content} & Categorised collection of topics & Reflective loop & Dedicated planning process and ``act'' step & Clarification and wording & Added guideline & Refined guideline and streamlined model \tabularnewline
\textbf{Validation} & & & Applied to 4 courses & Applied to 2 courses & Applied to 2 courses & Applied to 9 courses \tabularnewline
  \bottomrule
\end{tabular}}
\label{tab:iterations}
\end{table}

\paragraph{Iteration 1 and 2}
We based the first iteration of the model on a structured collection of post-it notes. In the initial collection of ideas, we categorised topics encountered during a joint brainstorming session in groups that included ``type of involvement'', ``constraints and forces'', as well as ``conflicts and challenges''. We limited the purpose to a pedagogical scope, since the decisions about the course should be driven by pedagogical considerations. Two of the teachers then used this topical collection to create an initial draft of the model that included a rough structure of a reflective loop (plan-act-reflect), with the different topics associated with certain phases of the loop.
We depicted constraints, conditions, actors, and forces in a circular dependency to indicate that these aspects influence each other strongly. To analyse the effects of decisions in the planning and acting stage on the course and derive meaning from them, we used Brookfield's four lenses~\cite{brookfield1995becoming}, which we extended to also include the university as a whole and the external stakeholder perspective. Once this meaning had been derived, we could  use it to reflect on the current state of the course, what the state of the course should be, and how to arrive there \cite{smith2001formative}.

\begin{figure}[tb]
\includegraphics[width=0.99\textwidth]{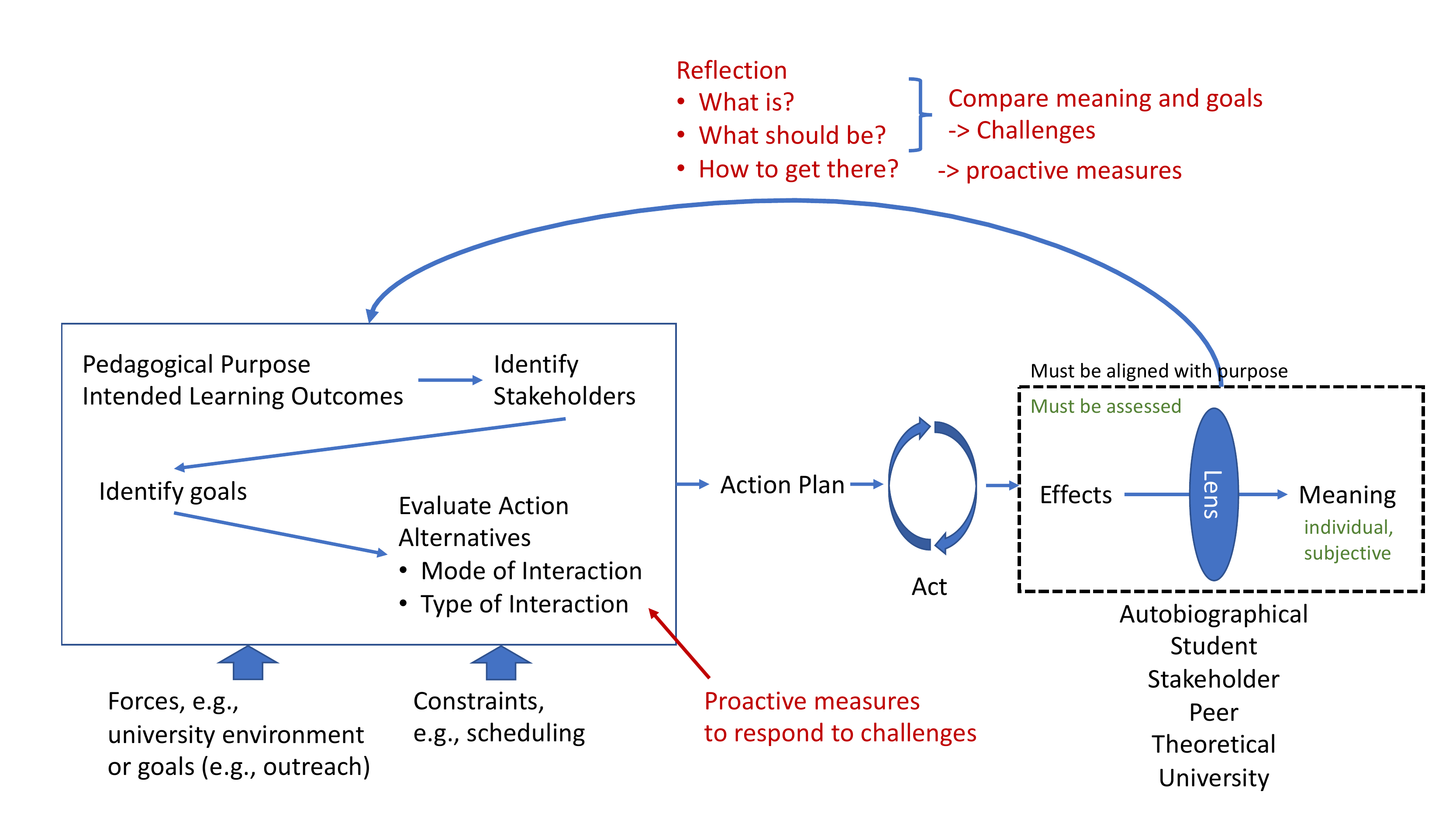}
\caption{The third iteration of the model with a dedicated planning process, an ``act'' step, and externalised forces and constraints.}
\label{fig:research-methodology:model:v3}
\end{figure}

\paragraph{Iteration 3}
The third iteration of the model was the result of a discussion of the previous iteration in the guild of teachers. The main topics of that discussion were a) the circular depiction of constraints, conditions, forces, and who, as well as b) the correct position of the purpose in the model. Furthermore, the group discussed the fact c) that a concrete step for acting on the actions was missing. To address issue a), we moved constraints and forces outside of the planning step, and depicted them as influences coming from the outside. This view is consistent with the fact that these constraints and forces are usually outside of the teacher's control, but must be adhered to in the planning process. In addition, we introduced an explicit process with discrete steps in the planning. The identification of stakeholders replaces the abstract ``who'' in the previous iteration. Identifying stakeholder goals becomes a distinct step that feeds into the evaluation of different action alternatives that finally yields an action plan. We addressed issue b) by moving the pedagogical purpose, now augmented with the intended learning outcomes, to the beginning of the reasoning process. This strengthens the view that all actions the teacher takes should be \emph{constructively aligned} \cite{biggs1996enhancing} with the intended learning outcomes. Finally, we introduced an explicit ``act'' step, thus addressing issue c). Additional changes include the separation of the ``peer'' and ``stakeholder'' lenses, the clarification of the first two steps of the reflection, and concretising the feedback from the reflection loop by clarifying that the identification of steps how to achieve the desired state is equivalent to defining reflective measures. 

This iteration of the model was the first one that some of the teachers in the guild applied to analyse their previous courses. In total, we applied the model to four courses. General issues that were reported were unclear definitions (e.g., the difference between ``type of interaction'' and ``mode of interaction''), unclear descriptions of activities (e.g., ``Identify goals'' was considered too generic), and the value of the ``act'' step.


\paragraph{Iteration 4}
Compared to Iteration 3, we made only small changes in this iteration based on feedback from the previous iteration. In essence, we made the following changes:
\begin{longitem}
	\item A ``context'' was introduced as an additional influence on the planning phase to capture aspects such as the course size and the student background that were not well described by ``forces'' and ``constraints''.
	\item The ``Identify goals'' step was replaced with ``Identify goals and trade-offs for each stakeholder'' to clarify this step.
	\item A ``risk assessment'' step was added to ``Evaluate Action Alternatives'' in the planning phase.
	\item The ``effects'' in the phase after the act step was replaced by ``observations'' to clarify that the effects will only become pertinent if observed by the teacher.
\end{longitem}
Two teachers attempted to apply this model to their courses. An important new challenge was insufficient guidance on how to apply the model. While we had reached a consensus about the meaning of the different aspects of the model, it was still difficult for teachers to consider all relevant aspects. 

\begin{figure}[t]
\includegraphics[width=0.99\textwidth]{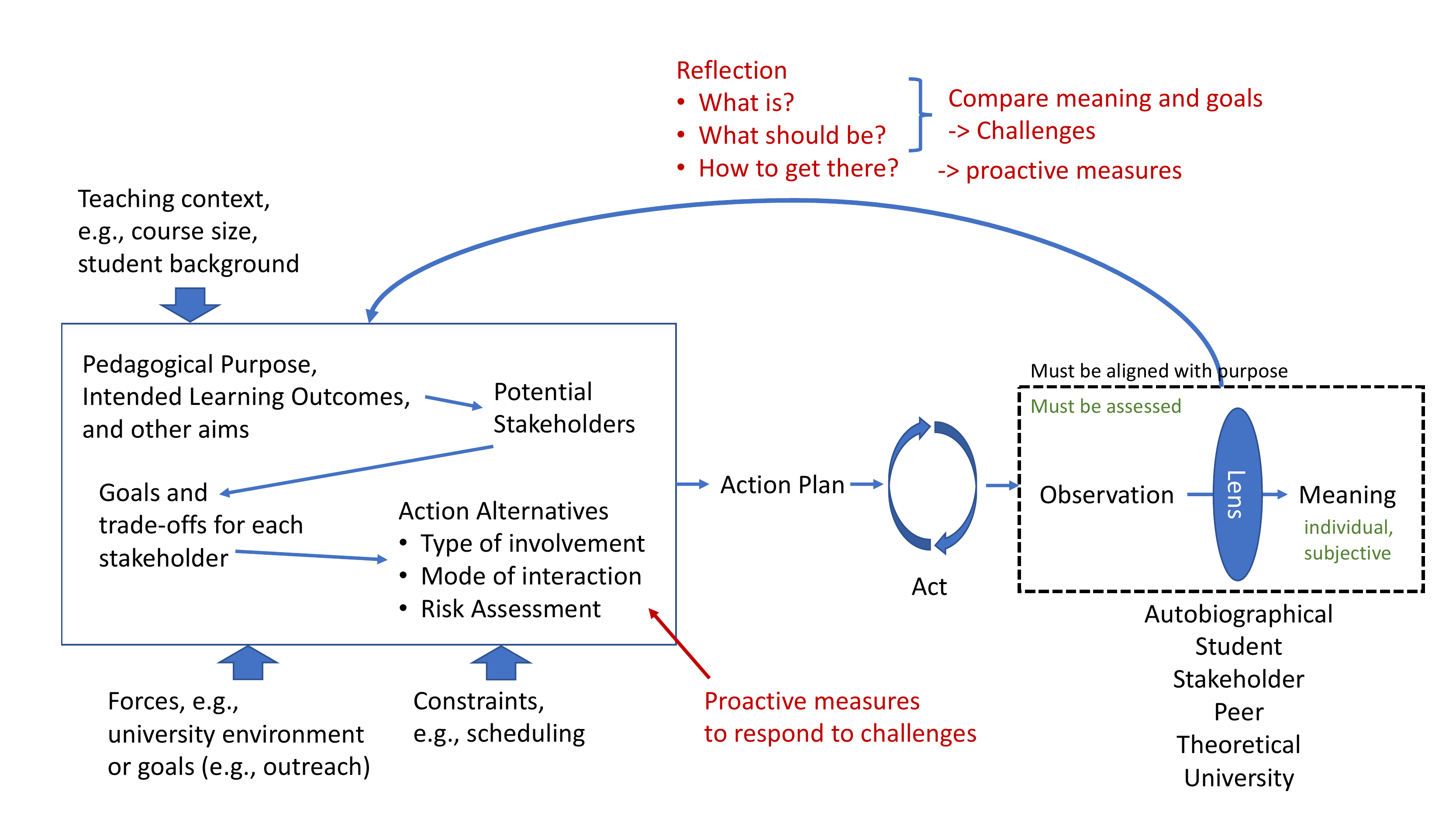}
\caption{The fifth iteration of the model with minor changes to the wording and structure.}
\label{fig:research-methodology:model:v5}
\end{figure}

\paragraph{Iteration 5}
While the next iteration of the model, as depicted in \fref{fig:research-methodology:model:v5}, saw only minimal changes in wording or structure, we added a ``guideline'', a set of questions that would allow a teacher to fill in the different parts of the model and check if all aspects were considered.
 The final iteration of the guideline is part of the appendix of this paper (cf.~\aref{sec:appendix:guideline}). The guideline is structured in five parts, corresponding to the different components of the model:

\begin{longitem}
	\item \emph{Part A: Influences:} This part contains questions about the teaching context, forces, and constraints. The answers to these questions thus define which influences the teacher is able to plan and execute the course under.
	\item \emph{Part B: Prepare stakeholder involvement:} Questions about pedagogical purpose, potential stakeholders, goals and trade-offs, as well as possible action alternatives are clustered in this part. It thus corresponds to the complex on which the influences act.
	\item \emph{Part C: Action Plan:} The definition of the action plan is placed in its own part since it is the main artefact created in the process. The questions in this part help the teacher identify which of the action alternatives should be applied and ensure that justifications for all choices are provided.
	\item \emph{Part D: Observation and Analysis:} Questions about the observation direct the teacher's attention to different aspects, such as student and stakeholder behaviour, and the teacher's own ability to reach her aims. The analysis part is addressed with questions that are aligned with the extended Brookfield's lenses, and help the teacher derive the meanings from the observations.
	\item \emph{Part E: Reflection:} Questions aimed at deriving the current status and the ideal status are contained in this part. From the comparison of the two, proactive measures should be distilled.
\end{longitem}

We applied this iteration of the model and the guideline to two courses. While the guideline provided good support, some questions were not clear and the teachers felt that some of them might not apply to their specific course setup. This triggered a discussion about the purpose of the model and the guideline. During this discussion, many possible purposes were discussed. Some of these provide valuable perspectives. For instance, the idea that the model and the guideline are tools to find conflicts between the different stakeholders in the course by allowing an analysis through the different lenses is certainly in line with our original intentions. Using the model and guideline as a way to communicate feasibility and expectations to external stakeholders might also be a possibility. The result of the discussion, however, focuses again on our main aim, which is to use the model as a tool to structure our thinking about involving external stakeholders in project courses, and the guideline as a description of the model and a ``data collection guide'' for collecting data about courses that already have taken place.

Additional remarks by the teachers were that guidelines are needed to apply the model to different situations. A project course in which external stakeholders act as product owners, e.g., is different from a course in which they act as guest lecturers. While we felt that the model covers both these situations, the guideline in its current form was too rigid and extensive to apply, especially in the latter situation. 


\paragraph{Iteration 6}
While we described the final iteration of the model and the guideline in \sref{sec:model}, the most important changes from iteration 5 to iteration 6 are described here.
We renamed the guideline to ``data collection guide'', to clarify its intention and the way it is supposed to be used. Additionally, we added examples to some questions to provide additional guidance and disambiguated the term ``aim''.

Most importantly, we changed the structure of the model as depicted in \fref{fig:conceptual-model}. The different parts of the plan-act-observe-reflect cycle have been condensed and the lenses are now applied to all aspects of the model, instead of just the observation. This emphasises the need to include the perspectives of all relevant stakeholders in all stages of the planning. We also changed the semantics of the boxes so that they are now activities and labelled arrows, with a noun describing what is carried from one activity to the next, or what influences the next activity.
We applied this final iteration of the model to several courses, as described below and in \sref{sec:application}.

\subsection{Data Extraction to Identify Common Risks, Aims, and Mitigation Strategies}
\label{sec:research-methodology:mitigation}

In order to extract data from our application of the models, the course coordinators (the teachers owning one course) filled in the guideline of model iteration 6 and gave the filled-in guideline to another teacher. The other teacher then used a small set of pre-defined codes (risk, mitigation strategy, success story, aim) to code the different aspects. Each coder also extended the codes as required. This way, a number of sub-codes were introduced, e.g., for specific risks or aims. We collected all these aspects in a shared spreadsheet and coders could cross-reference the codes of others. One round of feedback was built into the process. Each coder had the chance to speak to the course coordinator to clarify any points in the filled-in guideline. This proved to be a crucial step in the process and will later on be referred to as \emph{peer audit}.

We identified a total of 52 risks, 38 mitigation strategies, 17 success stories, and 25 aims in this initial coding. One of the researchers involved in the study coded the risks and identified a total of eight themes that showed up repeatedly and discussed them in a meeting with one other researcher. On reaching agreement about the meaning of the themes, the themes were assigned to the lens or to the axis between the lenses.
To do this, we worked with a model of the interactions of the relevant lenses involved in the courses (cf.~\fref{fig:results:risk-themes}). This model contains axes between relevant lenses. We excluded the program/university and the peer lens, since these interactions are not pronounced in the information we got from the model. Since it was us teachers who analysed our own courses, the main perspective is that of the teacher, while the main axis of interest is the one between the external stakeholder and the students. This axis is at the same time the most difficult to observe and influence, but also the one with the biggest impact on course success. 

Using this model helped the second researcher in assigning the risks to the eight themes. 
One of the original themes was eliminated and the researchers were able to refine the original assignment and the meaning of the themes. A third researcher was then asked to validate the assignment.
After the risks were assigned to a theme, all risks within the theme were reviewed in order to see which could be merged. 

Each involved researcher assigned the mitigation strategies they found in the course they coded to one or more risks, as reported in the model. After the final set of themes emerged, the mitigation strategies were assigned to the themes by using these risks as the guidance. That created a mapping of mitigation strategies to risk themes.

The final set of risk themes, of simplified risks, and of mitigation strategies was then reviewed by all researchers involved in the work.

\subsection{Threats to Validity}
\label{sec:research-methodology:threats-to-validity}

The model in \sref{sec:model} as well as risks and mitigation strategies reported in \sref{sec:results:risks-mitigations} are derived from our own discussions regarding the experiences of involving external stakeholders in our courses. To counter bias, each author coded the data supplied for a course in which they are not personally involved. Still, the spread of courses that were available to analyse and the background of the researchers is biased towards Swedish universities. While difficult to quantify from our data sources, there was a feeling that the Swedish universities had a more open attitude towards involving external stakeholders, which implies that other teachers and educational settings will have differing experiences than those we have. 

Two courses were coded by two authors independently of each other. This served as a sanity check. No disagreements in terms of semantics between the teachers were found, even if the same meaning could be annotated with different terms. So, even if the terminology of the codes differed, the intended meaning was the same. We also applied different lenses when collecting and analysing data, aiming for theory triangulation~\cite{Stake1995}. However, there was no involvement of external audits, including no involvement of external stakeholders in the coding process. It is possible that coders with another background and perspective would have found other interpretations of the data. The outcome of the analysis, both from applying the different iterations of the model in general and iteration six in particular, was discussed at multiple meetings in order to confirm the validity of the analysis \cite{Shenton2004}.

We count the number of risks and mitigation strategies that were grouped under each theme (cf.~\tref{tab:results:risk-themes}). These numbers reflect the proportion of reported risks and mitigation strategies and should not be interpreted as the severity or probability of a risk. Neither do they necessarily represent the relative importance among themes, either since different authors have focused on different aspects of involving external stakeholders and have differing experiences. 

The risks and mitigation themes are a data-driven and evidence-based description of the complexity of involving external stakeholders in project courses. We make no claim that the themes are exhaustive. As more teachers report on their experiences, the themes might be extended or new themes might emerge. Neither do we claim that the themes are orthogonal. The risks and mitigation strategies involve multiple actors and therefore represent more than one perspective on the collaboration. One risk can also be a cause for another risk. It is thus not always possible to define the cause and effect logic behind risks since there can be causation chains or cycles of risks. 



\section{A Conceptual Model of External Stakeholder Involvement}
\label{sec:model}


As a result of our discussions in the guild of teachers, we present a model of how to plan, act, observe, and reflect \cite{Kolb2014} on 
external stakeholder involvement in courses in academic education 
(cf.~\fref{fig:conceptual-model}). In order to keep the model's reflective step concrete, we follow Smith's definition of reflection: \emph{``assessment of what is in relation to what might or should be and includes feedback designed to reduce the gap''}~\cite{smith2001formative}.

The model is intended for use in three ways:
First, the model can be used \emph{retrospectively} to analyse a course that involved external stakeholders to help teachers identify sources of experienced frustrations and possible mitigation strategies. This allows a guided reflection of the teacher and enables a transfer of knowledge to future course instances.
Second, the model can be used \emph{constructively} to plan the next iteration of a course and future collaborations. Using the model this way, it is possible to build on knowledge from the retrospective application as well as from a general body of knowledge, e.g., about possible mitigation strategies, their context, and their effects.
Ideally, the uses would be combined and the model would be applied iteratively: the knowledge gained from using the model retrospectively would influence the constructive application for a new course instance that would in turn be analysed by applying the model retrospectively.
Finally, the model points to areas of future research and we supply our own data sources to be re-analysed or included in broader or more descriptive studies on the drivers and barriers for including external stakeholders in engineering education.
We have used the model in both the retrospective and the constructive ways, but mainly report on its retrospective application in the following. A brief overview of our experience in using it constructively is given in \sref{sec:constructive}.

\subsection{Structure of the Model}

The final iteration of the model is depicted in \fref{fig:conceptual-model}. It is accompanied by a guideline (cf.~\aref{sec:appendix:guideline}) that is comprised of questions that a teacher can answer when analysing a course. The model is split in three main steps, following the plan, act, and observe plus reflect cycle~\cite{Kolb2014}:

\begin{figure}[b]
\includegraphics[width=0.99\textwidth]{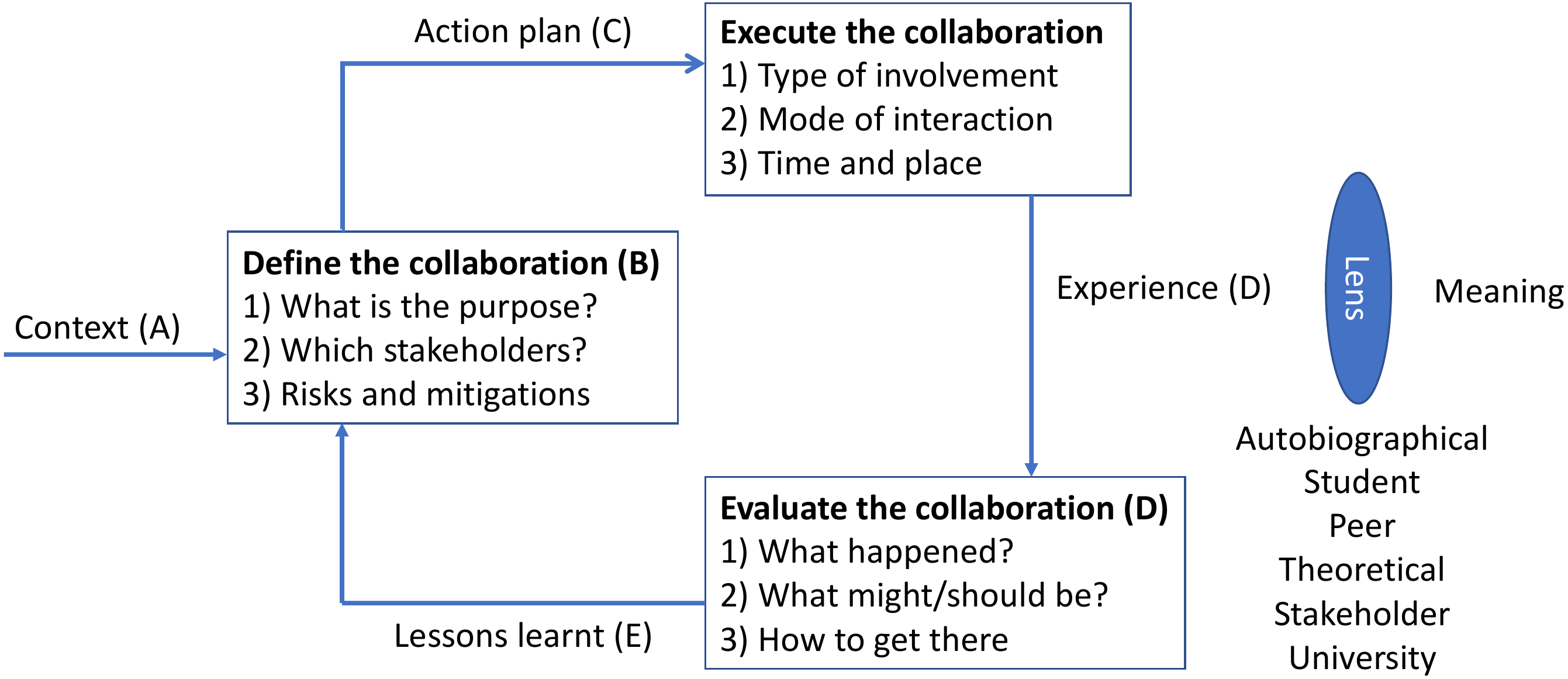}
\caption{A conceptual model of external stakeholder involvement. Letters and parentheses indicate the corresponding part of the data collection guide.}
\label{fig:conceptual-model}
\end{figure}

\begin{enumerate}
	\item Define the collaboration: plan the collaboration with the external stakeholder by defining goals, identifying potential external stakeholders, and risk as well as mitigation strategies.
	\item Execute the collaboration: work with the stakeholders through different types and modes of involvement and at the defined time and place.
	\item Evaluate the collaboration: based on the collected data, observe what has happened, how it differs from an optimal execution and outcome of the course, and identify strategies to improve the external stakeholder interaction in the future.
\end{enumerate}

The extension of Brookfield's lenses we use covers all of these steps. The main lens is the autobiographical lens, i.e., the one that focuses on the experiences of the teacher. Since the teacher is the one planning and conducting the course as well as the main interaction partner of both students and the external stakeholder, and ultimately carries the responsibility of the course, she will also be the one contributing most of the information to the planning and the observation. In addition, changes in the course will most likely be triggered by the teacher or at least executed by the teacher if reacting to an external stimulus. An overview of all lenses involved in the model, the data they can provide for its application, and the effect they have on the outcome of the model is provided in \tref{tab:model:lenses}.

\begin{table}[b]
\centering
\tbl{The different lenses and their impact on the application of the model. Lenses not included in Brookfield's original suggestion are highlighted.}{
\begin{tabular}{@{}P{1.6cm} P{3.5cm} P{3cm} P{4.4cm}@{}}
  \toprule
   \textbf{Lens} & \textbf{Explanation} & \textbf{Data Provided} & \textbf{Effects} \tabularnewline
  \midrule
Auto\-biographical & The teacher that applies the model. & Own observation, course notes, reflections & Responsible for all decisions in the course, main contact person of the students and the external stakeholder, held accountable by the university. \tabularnewline
Student & Students directly interact with external stakeholders and teachers; need to reach course's learning objectives. & Course evaluation, exam results, oral feedback & Are the main beneficiaries of a working external stakeholder involvement; abilities constrain stakeholder involvement \tabularnewline
Peer & Other teachers that engage in discussion or peer audits of the course. & Shared experience, can give hints about relevant literature or pedagogical concepts & Can play advisory role \tabularnewline
Theoretical & Information from published literature. & Can provide mitigation strategies or inform the pedagogic decisions of the teacher. & Allows argumentation for certain solution approaches, provides theoretical foundation for course development.  \tabularnewline
\midrule
\emph{External Stakeholder} & The concrete people involved in the course, representing a corporation & Feedback to the teacher, often orally & Goals and engagement define extend of involvement \tabularnewline
\emph{University} & The study program and university represent the broader educational context of the course. & Context such as schedule and limitations & Limits teacher's options but can also provide opportunities (e.g., by using established collaborations of study program)\tabularnewline
  \bottomrule
\end{tabular}}
\label{tab:model:lenses}
\end{table}

\subsection{Data Collection Guide}

The main purpose of the data collection guide is to reveal implicit assumptions, factors that influence the involvement of external stakeholders in a course, to allow the planning of the involvement, and to guide the reflection about the stakeholder involvement. As such, the guide follows the three steps outlined above. They correspond to Parts B, C, and D and E of the guide. Part A of the guide covers the context of the course and thus the arrow going into the cycle depicted in \fref{fig:conceptual-model}. Each part is split into more fine-grained sections that address specific aspects. For each aspect, a number of questions are defined. These questions do not necessarily all have to be answered --- instead, they are designed to cover many possible circumstances of a course and trigger reflection by the teacher. If a question is not applicable to the course, the teacher is encouraged to ignore it.

The questions in the guide are formulated in a forward-looking way. As such, they are formulated so as to apply mostly to the constructive use of the model. When applying the model retrospectively, their intention is still clear, however. One aspect that needs to be considered is that the difference between the planning and the concrete action plan is not entirely clear when applying the model retrospectively. Unless extensive documentation of the planning process of the course exist, most teachers answer the questions in Part B of the guide that are concerned with planning with the actual outcome they observed when \emph{acting}. This is not an issue per se, but makes the questions in Part C that is concerned with the action plan feel redundant.

Answering the questions in the guide help in addressing the three parts of the model. The context of the course is determined in Part A of the guide with questions such as: \emph{``Does the university have goals in place that require teachers to include external stakeholders? What do they mandate?''}. If the teacher does not know the answer to this question, it might trigger an exploratory process in which the teacher learns more about the teaching environment. Likewise, general course planning questions such as \emph{``Are there external events that need to be considered in the planning of the course?''} are part of the guide since they have a great influence on the possible stakeholder interactions.

Part B of the guide then addresses the definition of the collaboration. It contains questions about the pedagogical purpose and intended learning outcomes such as \emph{``Which, if any, learning objectives justify the inclusion of external stakeholders?''} and helps in identifying the capacity and capabilities of potential stakeholders with questions such as \emph{``Is there support by the management of the external stakeholder?''}. Aims and trade-offs for each stakeholder are also polled and serve as the input for the determination of action alternatives, including the type of interaction (which ``services'' can the external stakeholder provide?), the mode of interaction (how can students interact with the stakeholder?), and the alternatives' respective risks.

Once these alternatives are enumerated, Part C of the guide is concerned with identifying a viable \emph{action plan} by drawing on the knowledge about the stakeholders' schedules, type and mode of interaction, risks, and mitigation strategies. This corresponds to the input of the execution of the interaction.

Part D allows the teacher to structure her observations and assign meaning to them according to the different lenses. It thus corresponds to the transition between \emph{Execute the collaboration} and \emph{Evaluate the collaboration} in \fref{fig:conceptual-model}. The questions focus on behaviour and satisfaction of the students, the behaviour of the stakeholder, and the teacher's ability to teach. Alignment of the aims and interaction with other courses in the program are also polled. The assignment of meanings from the observations are strictly separated by lens and include questions like \emph{``Did the stakeholders gain value from their involvement?''} (stakeholder lens) and \emph{``How do the effects and the meaning you derived from it compare to reported accounts?''} (theoretical lens).

Reflection is guided with a set of questions in Part E of the guide. The meanings derived in Part D can be used to answer \emph{``What was?''} to determine the status quo. \emph{``What might or should be?''} can be inspired by the original aims and plans determined in Part B and C, or by a new understanding of the course that has developed while it ran. \emph{``How to get from what was to what should be?''} then allows the formulation of new mitigation strategies, new course objectives, new types or modes of interaction, or other new ideas that can improve involvement of external stakeholders in the future.

Applying the guide thoroughly yields a documentation of the course that is useful beyond the involvement of external stakeholders. It can be used to discuss many aspects of a course, such as problems with scheduling or prerequisites. However, the focus on stakeholder involvement allows a teacher to reflect deeply on this issue and explore the teaching environment for this specific topic. The way the model was applied to different project courses is detailed in \sref{sec:application}. The results of this application are detailed in \sref{sec:results}. An example of the constructive use of the model is given in \sref{sec:constructive}.

\section{Application of the Model to Project Courses}
\label{sec:application}

 \begin{table}[b]
\centering
\tbl{Overview of the courses that have been analysed retrospectively (cf.~\sref{sec:application}) and constructively (cf.~\sref{sec:constructive}). Student numbers are approximates per instance.  (CTH: Chalmers University of Technology, GU: University of Gothenburg, BTH: Blekinge Institute of Technology, BU: Bo\u gazi\c ci University, U of T: University of Toronto). The number of students refers to course instance; the number of stakeholders is in total for all instances.}{
\begin{tabular}{@{}P{4.2cm} P{1.5cm} p{1.4cm} p{1.4cm} p{1.3cm} p{1.8cm}@{}}
  \toprule
  Course & University & Level & Instances & Students & Stakeholders \\
  \midrule
  \multicolumn{6}{@{}l}{\textbf{Retrospective use of the model}} \tabularnewline
  \midrule
  CMP450: Software Engineering & BU & BSc & 1 & 70 & 6 \\
  CMP451: Project Development in Software Engineering & BU & BSc & 1 & 70 & 1 \\
    CSC302: Engineering Large Software Systems & U of T & BSc & 1 & 50 & 1 \\
  DAT255/DIT543: Software Engineering Project & CTH \& GU & BSc & 4 & 60/140 & ca. 20 \\
  DAT265/DIT599: Software Evolution Project & CTH \& GU & MSc & 2 & 50-60 & 9  \\
  DIT029: Software Architecture for Distributed Systems Project & GU & BSc & 1 & 70 & 1 \\
  PA243: Master Thesis in Software Engineering & BTH & MSc & 3 & 50 & ca. 30 \\
  TDA593/DIT945: Model-Driven Software Development Project & CTH \& GU & MSc \& BSc & 2 & 150 & 2 \\
  \midrule
  \multicolumn{6}{@{}l}{\textbf{Constructive use of the model}} \tabularnewline
  \midrule
  EDA397/DIT191: Agile Development Processes & CTH \& GU & MSc & 4 & 80  & N/A\\
  \bottomrule
\end{tabular}}
\label{tab:courses}
\end{table}






We applied the model both retrospectively and constructively to eight project courses (cf.~\tref{tab:courses}) in software engineering programs at the University of Gothenburg, Chalmers University of Technology, and Blekinge Institute of Technology in Sweden, the University of Toronto in Canada, and Bo\u gazi\c ci University in Turkey.
We put a particular focus on project courses and capstone projects since 
these often involve external stakeholders, present an environment in which stakeholder involvement can take extremely diverse forms, and where many different types of stakeholders can be involved. However, we have also included other courses with a project component in this analysis to show that our model is applicable in a variety of circumstances.

From our retrospective use of the model, we deduce a number of common challenges we see, as well as a selection of mitigation strategies that we have tried. A particular focus is put on the context in which these mitigation measures can be applied and which assumptions are made for each.
The constructive use of the model illustrates how external stakeholder involvement can be planned and constructively aligned with the learning objectives and the assessment. Each course is identified with its course code since some courses can have changed the course plan at the time of reading, through the code it should be possible to find the correct course plan from each university.


The rest of this section will highlight the most important aspects from applying the model to our own courses in terms of the \textit{roles} in which the external stakeholders participated, the \textit{observations} from applying the model, and the \textit{outcome} as lessons learned about the involvement and possible improvements. 
The full description of all eight courses course and model application experiences can be found in \aref{sec:appendix:coursedescriptions}.  
While this section provides an overview, these long-form descriptions can enable the reader to better assess the transferability of the types of external stakeholder involvement, success stories, risks, and mitigation strategies to their own educational aims and settings. 



\subsection{CMPE450: Software Engineering Course}
\label{sec:application:cmpe450}
\noindent \textit{Roles:} University employees act as customers. 

\noindent \textit{Observations:} The model is helpful in showing the current course design in a systematic way. It takes too long time to complete the final version of the guideline, and you need time to get the bigger picture after focusing on the guideline details.

\noindent \textit{Outcome:} A plan for the maintenance of the student software needs to be in place if the software is to be used in beyond the course. Therefore, involve the university's IT staff before preparing the course project description.
Customers might have unrealistic requirements. Monitor the students' requirements gathering and elicitation process and intervene if students get confused.   

\subsection{CMPE451: Project Development in Software Engineering}
\label{sec:application:cmpe451}
\noindent \textit{Roles:} Software practitioner and entrepreneur as course's primary teacher.

\noindent \textit{Observations:} The model helps to think beforehand who the right candidates for external stakeholders are. It also shows the stakeholders that might have been implicitly introduced in the course without proper awareness. The template needs to be filled out separately for each stakeholder with different roles and responsibilities. 

\noindent \textit{Outcome:} External stakeholders whose engagement cannot be guaranteed should not be included in the course, even when teachers can prevent unpleasant consequences by taking over those responsibilities.

\subsection{CSC302: Engineering Large Software Systems}
\label{sec:application:CSC302}
\noindent \textit{Roles:} Industrial guest lecturers.

\noindent \textit{Observations:} The model facilitated explicit focus on the variety of stakeholders and their goals, and also captured both known and previously unknown disconnects between important stakeholders. Since the guideline is formulated in a constructive way, many questions are redundant or not applicable for retrospective course analysis. 

\noindent \textit{Outcome:} External stakeholder use was misaligned to course content. More interestingly, the model brought to light issues with departmental support for guest lectures, and lack of governance. Teachers need departmental support to coordinate with guest lecturers, specifically sessional instructors who rely on others' contacts.

\subsection{DAT255/DIT543: Software Engineering Project}
\label{sec:application:dat255}

\noindent \textit{Roles:} Product owners, mentors, guest lecturers and jury members.

\noindent \textit{Observations:}
The questions are mostly formulated for constructive use, not for evaluation. Knowing the guideline in advance avoids answering questions twice. 

\noindent \textit{Outcome:}
By applying the model we identified different types of stakeholders but also issues unrelated to stakeholder involvement. We saw that there is little alignment of the course objectives with the educational program and no means to gauge whether the aims of all involved parties are achieved, which has led to discussions on program level to improve cohesion. The external stakeholders were insufficiently prepared for their role and the teachers' and stakeholders' aims were often mismatched. This will be handled by having clear statements of aims from all parties before the course starts. 

\subsection{DAT265/DIT599: Software Evolution Project}
\label{sec:application:dat265}
\noindent \textit{Roles:} Open source projects as product owners.

\noindent \textit{Observations:} It was easy to apply the model and it helped to document and cover aspects of the stakeholder interaction and set-up. Not all questions were relevant for all courses. A user needs to be selective when filling in the guide.

\noindent \textit{Outcome:} We found a misalignment between treatment of newcomers in open source projects and the time frame of the course, which led to adaptations of the assessment between two course instances to reduce the risk that students miss course goals by focussing too much on getting a contribution to the OSS project accepted.

\subsection{DIT029: Software Architecture for Distributed Systems}
\label{sec:application:dit029}
\noindent \textit{Roles:} Industrial company as product owners.

\noindent \textit{Observations:} The model helped identifying the challenges, risks, and mitigation strategies associated with involving an industrial company as product owner in a project course. When addressing a specific question in the model, say observed problems from the perspective of students, it is often difficult to separate between what we as teachers think and what students have actually reported. Ideally, the template should have been filled by different course stakeholders: teachers, teaching assistants, students, and external company.

\noindent\textit{Outcome:} Due to time constraints, the teaching assistants were used as intermediates so that the students did not interact directly with the product owners. This caused confusion about priorities and motivations for certain requirements. We will apply the model constructively to communicate the experiences of the first instance of the course to the different stakeholders, highlighting both the success stories and observed problems as well as evaluate the experiences using the different elements of the model.

\subsection{PA243: Master Thesis in Software Engineering}
\label{sec:application:PA243}

\noindent \textit{Roles:} Industrial contacts for master thesis projects  

\noindent \textit{Observations:} Version three of the model allowed for more creativity in the analysis while version six allowed for a detailed analysis of the different stakeholders. The result from the earlier model was harder to compare with results from other courses while the final model took more time to fill out. 

\noindent \textit{Outcome:} The model exposed why certain changes to the course worked and others did not, and for whom. We also saw that it is worth knowing the roles of external stakeholders to better assess changes to a course. One identified problem is that some projects are initiated by one contact person but carried out by someone else at the company, often resulting in lack of mandates, changing priorities and/or new project foci.

\subsection{TDA593/DIT945: Model-Driven Software Development}
 \label{sec:application:tda593}
\noindent \textit{Roles:} Industrial guest lecturers. 

\noindent \textit{Observations:} It was easy to apply the model and it helped to understand in depth the stakeholder aims. This uncovered potential sources of conflict and/or friction between the stakeholders, and the course responsibles or students. Differences between some questions were not clear or felt repetitive. 

\noindent \textit{Outcome:} A misalignment between student background and the chosen abstraction level of guest lectures,  as well as between the guest lecture content and the remainder of the course caused confusion. We plan to have discussion sessions with the teachers in the end of each guest lecture, as well as to spend more effort to prepare guest lecturers before their lecture.

\section{Results of Applying the Model}
\label{sec:results}

Using data obtained from a total of 15 instances of eight project courses we have taught, we demonstrate the usefulness of applying the model retrospectively. Based on this use, we show that the challenges encountered in a diverse range of settings can be reflected upon and interpreted with the help of the model. Our empirical findings also show that a majority of the experienced challenges are common amongst different courses or at least share common roots. We validate the constructive way of working with the model by showing how we use it to plan future instances of project courses and how the model supports reasoning about alternatives and formulating a concise action plan that takes context, forces, and constraints into account while ensuring constructive alignment.

\subsection{Observed Risks and Mitigation Strategies}
\label{sec:results:risks-mitigations}

\begin{figure}[tb]
\includegraphics[width=0.99\textwidth]{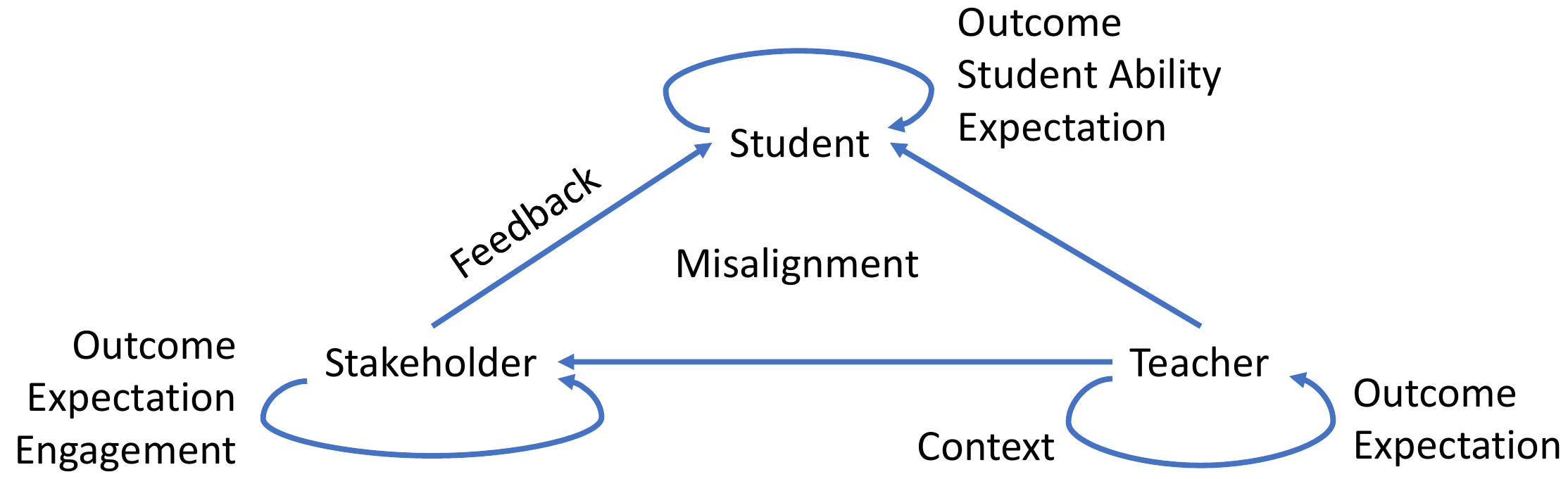}
\caption{The identified risk themes and their assignment to lenses or interactions between lenses.}
\label{fig:results:risk-themes}
\end{figure}

The data extraction described in \sref{sec:research-methodology:mitigation} led to a total of seven recurring risk themes. These themes are detailed in \tref{tab:results:risk-themes} and assigned to the different lenses and the interactions between them in \fref{fig:results:risk-themes}. The assignment was done by analysing where the risk originates, so that risks associated with student capability stem from the students' skill sets while contextual risks have their source in the teacher's planning in relation to overall educational constraints. All observed risks were assigned to one of the themes. Likewise, all mitigation strategies that were observed were assigned to the theme they address. A more detailed description of the risks and applicable mitigation strategies can be found in \sref{sec:results:lessons}.

\begin{table}[b]
\centering
\tbl{Risk themes, associated lenses, and their explanations.}{
\begin{tabular}{@{}P{.16\textwidth}P{.22\textwidth}P{.54\textwidth}@{}}
\toprule
Risk Theme & Lens & Explanation \tabularnewline
\midrule
Student Ability & Student & Risks associated with the knowledge, skills, and abilities of the students. \tabularnewline
Outcome & Student, External Stakeholder, Teacher & Risks associated with the desired results for all lenses. This includes grades, course evaluations, stakeholder goals, and others. \tabularnewline
Expectation & Student, External Stakeholder, Teacher & Risks associated with the aims and motivations of the different lenses w.r.t.~the course and the interactions. \tabularnewline
Engagement & External Stakeholder & Risks associated with the effort, time, or resources the external stakeholder can invest in the course. \tabularnewline
Context & Teacher & Risks associated with university or study program related issues that are beyond the teacher's control. \tabularnewline
Feedback & Between External Stakeholder and Student & Risks associated with the interaction between the external stakeholder and the students. This interaction is crucial for the success of the course but hard for the teacher to observe and control. \tabularnewline
Misalignment & Between all lenses & All risks that are associated with integrating themes across lenses. For instance, while the expectations of the external stakeholders and of the teachers on their own can be perfectly reasonable, compared to each other they can be misaligned. \tabularnewline
\bottomrule
\end{tabular}}
\label{tab:results:risk-themes}
\end{table}

Mitigation strategies are used to gain control over a risk factor. We have identified mitigation strategies for all but one risk theme: there is no mitigation strategy for context (apart from planning accordingly), since this is not under the teacher's control.
It is worth noting that the mitigation strategy for a risk can affect a different risk theme than the risk it addresses. For instance the risk \emph{``unlikely that external stakeholder can attend all lectures''} is assigned to the risk theme \emph{engagement}, but one of the applicable mitigation strategies, \emph{``external stakeholder has bi-weekly meetings with student groups''}, is assigned to the risk theme \emph{feedback}.
\tref{tab:results:themes-frequency} provides an overview of the themes and there frequency as risks and mitigation strategies with examples from our collected data. 

\begin{table}[b]
\centering
\tbl{Number of risks and mitigation strategies per theme and examples for risks and mitigation strategies for each theme.}{
\begin{tabular}{@{}P{1.7cm}P{1cm}P{0.8cm}P{4.4cm}P{4.2cm}@{}}
\toprule
\textbf{Theme} & \textbf{\# risks} & \textbf{\# mit.\ strat.} & \textbf{Example risk} & \textbf{Example mitigation\newline strategy}\tabularnewline
\midrule
Student\newline Ability & 5 & 2 & Students can't always transfer the knowledge of the external stakeholder to their own context. & Supervise students to help understand stakeholder perspective. \tabularnewline
Outcome & 7 & 6 &  Teachers' disappointment -- integrating stakeholder not worth the effort & Follow-up lecture where concepts are explained in relation to the remainder of the course.\tabularnewline
Expectation & 2 & 6 & Stakeholder does not see software development perspective and comes with unrealistic requirements. & Prepare stakeholder to understand student context.\tabularnewline
Engagement & 10 & 6 & External stakeholder has insufficient mandate/resources from management. & Ensure all involved external stakeholders have their managers/company permission and encouragement to participate. \tabularnewline
Context & 5 & 0 & Deadlines set by the university of program can not be adapted to the external stakeholder & \tabularnewline
Feedback & 8 & 8 & There is little control from academia over the quality of feedback from industry. & Regularly monitor interaction between student and external stakeholder\tabularnewline
Misalignment & 13 & 2 & Students are confused since external stakeholder does not stick to course goals or the course plan. & Align goals of the different parties and the course. \tabularnewline
\bottomrule
\end{tabular}}
\label{tab:results:themes-frequency}
\end{table}

\subsection{Lessons Learned about Involving External Stakeholders from Applying the Conceptual Model}
\label{sec:results:lessons}

With the risk themes in mind, we are able to provide an overview of the lessons we have learned about our courses from applying the conceptual model. The reflection that has been triggered by using the model unveiled several important aspects that were not considered before. These results can help teachers in structuring and planning the involvement of external stakeholders in their courses. While not all lessons learned are applicable to all types of projects and all types of interaction, there are some underlying principles that are generic enough to be considered in most circumstances. The lessons mentioned here are derived from our data collection and from the coding of the filled-in guides. We summarise risks in the following and connect them to the individual experiences of the teachers as reported in \sref{sec:application} and \aref{sec:appendix:coursedescriptions}, referring to specific courses when appropriate, and using the risk themes to structure the content.

\paragraph{Student Ability}
There are three major risks observed regarding student ability: that students are not able to transfer the knowledge of the external stakeholder to the context of their project; that the student skills and prerequisites are too heterogeneous to allow them to interact with the stakeholders on the same level; and that working with open problems, tools, processes, and external stakeholders can cause a cognitive overload among the students. All three risks are exacerbated by the fact that it is unknown whether students had previous interactions with stakeholders and possess the necessary skills to interact with the stakeholders effectively. In addition, the fluctuating workload of the students over the course of the term can have an influence on the effectiveness of the interaction.

To mitigate the risks concerning the students' ability the teacher can organise supervision to both help the students in how to use new processes and tools as well as in transferring the external stakeholder's perspective to the course's learning objectives and the project constraints.

\paragraph{Outcome}
Outcome is one of the risk themes associated with the student, the teacher, and the external stakeholder. Each of these lenses has a specific idea about the outcome of the project course. Some risks associated with outcome are cross-cutting. The fact that a plan of what to do with the products that are developed within a project after the course ends can be a risk for the student (who do not see their effort valued beyond the grade), the stakeholder (who has invested time and resources into the project but throws the end product away), and the teacher (who misses the chance to establish a learning sequence or exploit teachable moments about, e.g., software maintenance). An example for a lack of such a plan can be found in \sref{sec:application:cmpe450}.

Other outcome risks are more specific: if the intended learning outcome is not reached, the students suffer; if the course evaluation is poor because the external stakeholder did not meet the student needs, the outcome for the teacher is negative; if the external stakeholder does not reach their specific aims, the outcome for them is not worth the time and effort. If these risks manifest, the involvement of external stakeholders as a whole might be in question.

On the other hand, one of the risks associated with this theme is that it is difficult to measure or determine if the indented outcome has been achieved. Learning objectives should be tested in the course assessment, but it is more difficult for the external stakeholders and the teachers to identify if they have achieved their aims. As discussed in \sref{sec:results:strengths-weaknesses}, the conceptual model we propose can actually help in this regard. An explicit evaluation of the stakeholder interaction from the stakeholder's perspective as described in \sref{sec:application:PA243} can also support such a reflection.


These risks can be addressed in various ways. One way is to make the contradiction between different expectations a feature of the course so the students explicitly have to balance the outcome to fit the different stakes at play. Another strategy is to ensure that the involved parties have a more realistic view of the results. The students can be coached in how to transfer the stakeholder perspective to their own (see Student Ability). The external stakeholder can also be briefed before the collaboration begins to have a better understanding of what to expect. In both cases a systematic evaluation of the different expectations and perceived outcomes of the students and external stakeholders can facilitate a better understanding of their perspectives and help in conducting future course iterations.

\paragraph{Expectation}
Just like outcome, expectation is also a cross-cutting risk theme that is relevant for each of the three main lenses. However, our coding only revealed two explicit risks that we could clearly assign to that category. One reason for this is the fact that in many cases the problem is a \emph{misalignment} of expectations between different lenses, i.e., two expectations from, e.g., the stakeholders and the students, that in themselves are sensible, but are not compatible with each other. 

There are two risks that can be identified in the area and associated to one of the lenses. The first one concerns the students: they have an expectation that the product they produce (and put effort in) yields some benefit for them. This can, however, be stifled by an unresponsive stakeholder or by the aims they set for their product. Both issues can cause the students to work on a specific outcome for a long time, but their expectation of reward does not come to fruition. The second risk concerns the external stakeholder: especially in cases in which the external stakeholder has little experience with software development projects, their requirements might be unrealistic for the purposes of a student project (cf.~\sref{sec:application:cmpe450}). It is the responsibility of the teacher to adjust this expectation in advance, so as to avoid frustrations on the part of the student over an unrealistic project scope and on part of the stakeholder over not reaching the aims. 

\paragraph{Engagement}
Engagement focuses on the investment of the stakeholder in the course. Typical risks associated with this theme address a lack of resources or a missing mandate on the part of the external stakeholder. Such a lack of engagement often manifests by stakeholders that cancel meetings with students at short notice or are unprepared for their role (cf.~\sref{sec:application:dat255} and \sref{sec:application:tda593}). In the worst case, a lack of mandate can even mean that the external stakeholder can not participate in the course as promised, be replaced by another person, or is even forced to drop out of the course entirely. Such a situation can jeopardise the achievement of the learning objectives. For instance, in the master thesis project at BTH (cf.~\sref{sec:application:PA243}), the people that originally defined some of the thesis projects were no longer available as industrial advisors and were replaced with others who did not have a stake in the project or lacked experience or competence. This can cause situations in which the study is poorly designed or promised data cannot be made available. Since the teacher might not have much control over the external stakeholder, this can cause severe problems for the students and they might even be blamed for a sub-par outcome.

Personal differences, both in style and in approach, can also cause conflicts with the teacher or with the student. In some courses, we observed that stakeholders deviated from their briefing (cf.~\sref{sec:application:dat255}). This can confuse students and make it necessary for the teacher to step in. Related to this, involving external stakeholders in a course always bears the risk of a ``hidden agenda'' on the part of the external stakeholder. One of the reasons that the stakeholder deviated from the briefing was the deeply held personal belief that Scrum was not the best software development methodology. The stakeholder therefore communicated to the students that they should not apply it, even though this was the focus of the course. This is not an alignment concern since the risk originates by the external stakeholder's attitude and motivation to the other parties.

Generally, stakeholders need to be more aware of the students' background and needs as well as the course contents. Extensive briefings and discussions between the external stakeholder and the teacher are therefore unavoidable. For specific modes of interaction, the teacher can help align what was communicated by the stakeholder with the course content. A follow-up lecture by the instructor could, e.g., follow guest lectures to better align lecture content to course content.

\paragraph{Context}
Risks associated with the context are beyond the control of the teacher. Typical examples are deadlines set by the university (e.g., for examinations) and the study program (e.g., to report grades) that might conflict with the schedule of the external stakeholder. However, additional issues can appear when the project or course is regarded in the context of the study program. Prerequisites of the students, e.g., are usually acquired through other courses in the same program but might limit the possible types of projects that can be done. A lack of teaching sequences was also reported (cf.~\sref{sec:application:dat255}), indicating that there is poor alignment of learning objectives and teaching methods between the course and other courses in different programs. Insufficient guidance or lack of a strategy for involving external stakeholders can also be a problem (cf.~\sref{sec:application:CSC302}). Such issues can only be resolved on the level of the study program, however, and are somewhat independent of the involvement of external stakeholders. To know is a first step to be prepared.

\paragraph{Feedback}

The dynamics between the stakeholder and the students are the source of a number of risks. One main source of risks is the fact that the teacher struggles in monitoring and managing these interactions. On the one hand, the teacher has little ability or interest in controlling everything and imposing herself on the interaction. On the other hand, the fact that it is unclear what has been discussed between the students and the external stakeholder makes it more difficult to control whether learning objectives are achieved and whether the course content is delivered in a coherent fashion. A particular risk in this regard may be that the external stakeholder could give different groups of students different advise, or spend more or less time with them. That can create an unfair situation where some students receive ``preferred treatment''. If students rely too much on stakeholder feedback, in particular during the development process, they might focus too much on fulfilling stakeholder needs and not on achieving the learning objectives. 

If teaching assistants are involved in a course, there is a possibility that they act as a layer of separation between the external stakeholders and the students. In one of the courses (cf.~\sref{sec:application:dit029}) investigated, the students communicated their questions to the TAs who in turn relayed it to the stakeholders. This delayed the feedback and removed the interaction between stakeholders and students entirely, thus not achieving an important aim of the stakeholder involvement.

Depending on the students' and the stakeholder's personality, it might even be the case that students do not dare to approach the stakeholder and ask for feedback. This might cause a situation where the students are unable to fulfil stakeholder needs because they are unknown or where stakeholders feel that they are unable to reach their aims since the students are not in touch with them even though the offer has been made. Such an issue was, e.g., reported in \sref{sec:application:dat265}, where students did not approach the maintainers of the open source projects they were supposed to extend.

These risks can again be mitigated by the teacher stepping in and giving additional supervision, in this case to explain and reflect on the external stakeholder perspective. Another strategy is to be present and intervene when the students meet the external stakeholders and/or to talk with both parties throughout the collaboration. This can be managed by scheduling regular meetings and workshops. Yet another strategy is to encourage the students to communicate with the stakeholder outside of the scheduled interaction to receive the amount of interaction necessary to the project. However, this might trigger new feedback risks as the students are given conflicting advice or treated unfairly. 

\paragraph{Misalignment}

A recurring risk was that the expectations of the stakeholders w.r.t.\ the capabilities and the capacity of the students exceeded what the students were actually able to provide. Likewise, students often expected clear-cut answers and solutions from stakeholders who could only deliver nuanced descriptions. This latter issue was particularly observed for guest lectures which are often not appreciated fully by the students since they rarely provide directly actionable input and are often tangential to the course content. We refer to these risks as misalignment since the risk does not emanate from a specific actor of a collaboration but the discrepancy between the actors.

One particularly interesting case was observed in one of the analysed courses (cf.~\sref{sec:application:dit029}). The company involved in the course announced at the beginning of the term that they were recruiting and wanted to use the opportunity to see who of the students are potential candidates for hiring. Instead of feeling motivated, however, the students perceived this as a source of pressure and even mentioned that they felt exploited by the company. This is a case where the company had the best intentions but the expectation was misaligned with the ability and willingness of the students to deliver. In particular, this is in contrast to reports from the literature. Harrison~\citeyear{harrison1997enhancing}, e.g., states that employment prospects led to increased student motivation.

In another course in which students were asked to contribute to an open-source project (cf.~\sref{sec:application:dat265}), an inherent misalignment between the way open-source communities welcome newcomers gradually and the students' goal to make a significant contribution in one semester was evident. Open-source projects encourage new contributors to start out with simple code fixes, providing tests, and discussing issues on message boards and in bug tracking systems. The students, however, were asked to contribute a new feature or new functionality. This misalignment of expectations between all three parties should be kept in mind when using open-source communities as external stakeholders.

A major risk that can be observed in thesis projects (cf.~\sref{sec:application:PA243}) is that industrial partners are mostly interested in the product whereas the teacher is interested in approaching the topic as a research problem. This form of misalignment can cause the student to fail since theses are often evaluated based on their scientific contributions. A clear communication between the teacher and the external stakeholder before the thesis project starts needs to be in place to address this potential misalignment of expectations. Similarly, non-disclosure agreements that prevent a thesis from being published can prevent some learning objectives to be reached and are therefore a symptom of a misalignment between stakeholder expectations and desired outcome for the student.

Generally, student and stakeholder feedback should be collected throughout the course. This makes it possible to look for misalignment early on to take preventative actions. However, in many cases there are no explicit steps taken by the teacher to ensure that this feedback is collected in a timely and sufficient fashion. An exception is the master thesis project at BTH (cf.~\sref{sec:application:PA243}) where stakeholder feedback is explicitly requested, even though only at the end of the project. However, even at that point in time, the collection of feedback contributes to the organisational knowledge of the faculty and can be used in planning and executing future iterations of the project course.


\paragraph{Relation to Theory}
The majority of the courses we observed have different settings than the courses in the literature. In our cases, the teacher acts more as a facilitator. Much of the literature regards the teacher as more involved in the problem space and adheres to the guided instruction philosophy. Instead, the open problems tackled in the courses that are observed in our data collection are amenable to having an external stakeholder represent the problem experts. The level of involvement thus seems to differ from the majority of the instances reported in the literature, where the external stakeholder sponsors a case and acts as pure customer, as in the case of the Munich Airport~\cite{bruegge2008experiment} and many others, e.g., \cite{gabrysiak2012teaching,penzenstadler2014using,Boehm1998}. This is not only on the level of technical support, but also about communicating and collaborating between students and external stakeholders. 

Student ability is mentioned as a risk by Bruegge~et~al.~\cite{bruegge2015software} who report that students would add last-minute features just before the acceptance test which led to stability issues in the software. Similar scoping issues are reported by Penzenstadler~et~al.\ who saw that the students struggled to balance delivering what was needed to pass the course with what was required by the external stakeholders \cite{penzenstadler2013university}. More guidance to help the students handle the demands of the course and the external stakeholder as well as supervision regarding process aspects could have mitigated these risks. 

In terms of reported issues for engagement risks Penzenstadler~et~al.\ and Gabrysiak~et~al.\ found that the external stakeholders struggled in handling supervision for their student numbers \cite{penzenstadler2013university,gabrysiak2012teaching}, while Boehm~et~al.\ found that the external stakeholder was not prepared to give up-front assurance that they would use the outcome of the collaboration \cite{Boehm1998}. The consequence was a lack of mandate and resources to maintain the outcome. 

Penzenstadler~et~al.\  also report that the results did not fit the industrial requirements in their collaboration \cite{penzenstadler2013university}. The reason in their case was that the time frame did not allow for the complete verification and validation steps that were mandated internally by the external stakeholder. Similar issues are reported on by Kornecki~et~al.\ \cite{kornecki1997strengthening} and by Callele and Makaroff (cited in \cite{gabrysiak2012teaching}). This resonates with scheduling as a contextual risk factor and is not something that a teacher can influence due to the static deadlines imposed in academic education. 

Mitigating feedback risks is often a question of the teacher's availability. Kornecki~et~al.\ state that it can be difficult for teachers to invest as much time and effort as needed to fully understand the industrial case \cite{kornecki1997strengthening}, while Gabrysiak~et~al.\ experienced that it was not easy for students to get access to the external stakeholder outside of the scheduled interaction \cite{gabrysiak2012teaching}. 

Misalignment was reported by Callele and Makaroff (cited in \cite{gabrysiak2012teaching}) as a mismatch between the needs of the external stakeholder and the learning objectives of the course. This is not necessarily a problem if an explicit trade can be negotiated, such as providing time and effort in interacting with the students in terms of receiving innovative prototypes or new perspectives on old challenges (cf.~\aref{sec:appendix:dat255}). Another reported misalignment risk is that the students got worried about what to expect as exam questions in relation to the project scope \cite{daun2014industrial}. Again, supervision  and clarification from the teacher could have mitigated the risk given that there was time for such interventions. 

\subsection{Strengths and Weaknesses of the Model}
\label{sec:results:strengths-weaknesses}

While we explored the lessons learned about our courses in the previous section, we will describe our experiences with the model itself below. Since a total of eight teachers applied different iterations of the model to a variety of courses with different characteristics, we believe that our experience is broad enough to provide a good overview of the strengths and weaknesses. The fact that we are such a large group also counteracts the natural bias that we might have towards a model of our own creation --- while all teachers were involved in defining the model at one point, two of the teachers were driving its development.

We have gathered the strengths and weaknesses of the model from the individual reports of the teachers, found in an abbreviated form in \sref{sec:application} and in a longer form in \aref{sec:appendix:coursedescriptions}. They have been grouped according to the main aspect they address: data collection, perspective, scalability, understanding the courses, and model structure. In the following, we will discuss these strengths and weaknesses and derive recommendations on how to apply the model wherever suitable. These recommendations are also summarised in the instructions part of the guide in \aref{sec:appendix:guideline:instructions}.

\paragraph{Data Collection}
Working with the model and filling in the guide, in particular when used retrospectively, requires the teacher to draw from a rich pool of data about the course. However, it can be very difficult to evaluate the different lenses objectively since there is insufficient data available and little to no standardisation and formalisation to get this data. Even formalised data collection methods such as course evaluations suffer from a lack of participation. It is also very difficult to follow up on anonymous free-text feedback. This weakness can be somewhat addressed by extensive note-taking and follow-up interviews with students or external stakeholders, and potentially a peer review of the course. However, this effort will fall on the teacher and implies an additional workload. If a course is planned with the model, we encourage including data collection methods in the action plan.
At the very least, the model helps the teachers to understand this limitation more clearly and can spark further data collection, as has happened when we started our own reflection on our practice. Likewise, the model revealed that there was currently no way to measure the objectives of the different stakeholders and thus no way to gauge achievement, an important insight that can lead to further improvements of the course.

Another possibility would be to include students and the external stakeholder in the application of the model and have them provide the required information directly. This can be done in the form of a post-mortem as a group or by using the questions in the guide as an interview guide.

\paragraph{Perspective}
When applying the model, some teachers observed that the perspectives in the analysis part were not always clear. The way the questions are phrased motivates a focus on the teacher perspective. Due to this fact, it is not clear which risks revealed by the model are most critical to students. 
This issue is related to the data collection issue discussed above. Since insufficient data from the student and external stakeholder perspective is available, the teacher tends to automatically emphasise the teacher's perspective. This can be counteracted somewhat by employing data collection instruments that yield data from the other lenses. 

At the same time, the teacher mediates the interaction between the stakeholder and the student. The axis between stakeholder and student is not only a possible origin for risks, it is also the case that the teacher is rarely involved in this axis. The crucial difference between the involved lenses is that the teacher is the one who carries the responsibility. While there are complicated interactions between the different axes and who influences whom, the pattern that emerged from our analysis is focused on the teacher's perspective on the interaction between the external stakeholder and the students.

Even though there are limitations of the visibility of issues experienced by groups other than the teacher, the model and guide help consider the course and stakeholder involvement from many more perspectives than is often the case without them. In particular, the model also helps to see issues on a larger scale, e.g., on the scale of the study program, the department, or the university and can thus reveal issues in course and program governance. The insight generated in the application of the model can facilitate a dialogue with the program managers and department heads. We have observed that the model can act as a catalyst that can start extremely valuable discussions and help teachers to start thinking outside the box, in particular when it comes to the integration of the course into the study program(s) or the university strategy.

On a similar note, the model can help provide a clearer view of course issues to other academic stakeholders, helping to provoke understanding and change. This can be particularly helpful if the course is taken over by another faculty member. The rich data available through an application of the model makes many of the implicit assumptions explicit and makes the reasoning and the considered alternatives transparent. That saves the new teacher from repeating mistakes of the past. Thus, the filled-in guide can act as a communication instrument and as part of the organisational knowledge of the study program or department.

A peer audit with other faculty members not involved in the course helped in identifying the core issues with the courses and where implicit assumptions remained. Benefits include that there is a common language established between different teachers. These teachers do not necessarily have to be in the same organisation since our model supports the exchange of experiences across organisational boundaries.

\paragraph{Scalability}
In some of the courses (e.g., DAT255 described in \sref{sec:application:dat255} or the thesis course at BTH detailed in \sref{sec:application:PA243}), several stakeholders are involved. Their type and mode of interaction as well as their goals differ. Some act as mentors for the students, some act as case providers, and others as guest lecturers. In case of the thesis projects, stakeholder aims, motivations, and engagement, can differ for each individual thesis. When the model is applied to such a course, the effort increases significantly. If the guide is applied in full detail, the time required multiplies. A less detailed and more flexible iteration of the model might help alleviate this as well as instructions to use the model and guide more adaptively. The development and evaluation of such a iteration is, however, left as future work.

Furthermore, due to its length and the structure, it is difficult to extract valuable data from the filled-in guide. This is particularly the case when the model is applied in retrospective since the focus on details makes it hard to assess the big picture. Applying the coding procedure where the answers were annotated with risk themes can help in the data extraction process. However, several authors felt that version 3 of the model did indeed reveal more interesting findings despite its lack of clarity and the lack of a structured guide. This was specifically the case for courses in which the only mode of interaction were guest lectures.

Both versions of the model helped us better understand and analyse how the external stakeholders were used and involved in the courses. In relation to PA243, a master thesis project course, the final version of the model had more detailed questions and thus supported a more detailed analysis of certain aspects of stakeholder involvement. However, the large variation of thesis projects meant that it was perceived as more `costly' to apply this iteration to the many different stakeholder types. The third, and more `loosely' structured version of the model allowed for more creativity in the analysis. The result of this analysis was harder to compare to results from other courses and requires more background knowledge from the analyst. There is thus a trade-off between a detailed and more structure model, which ensures more comparable and detailed results, and a less detailed and unstructured model, which allows for more analysis freedom but also requires more detailed contextual knowledge. To cater for such different analysis situations the model might need to come in different versions, one being a less detailed iteration of the `full' one.

\paragraph{Understanding the courses}
Several teachers report that applying the model revealed a clear mandate for external stakeholders to be involved in and devote resources to the course. At the same time, when used retrospectively, it helped to better understand misalignment between course content and industry lectures. The retrospective use also helped in understanding how stakeholder involvement evolved over several instances and why past changes to the way external stakeholders are involved did or did not work.

An important strength of the model is its ability to make risks explicit. This in turn is a prerequisite for systematically tackling them and improving a course.
On the other hand, this risk analysis is not complete. For instance, the risk that a stakeholder won't come back the next year was not visible in the data. In general, we were looking mostly at the student perspective and not so much at the other lenses. When we compared our findings to the literature, we saw that some of the literature reports problems that exist in our courses, but that we did not consider when collecting the data. This emphasises the importance of the theoretical lens.

One weakness in applying the model as a teacher is that some issues are not visible because they are covered by other people. For instance, the teachers coding DAT255 (cf.~\sref{sec:application:dat255}) did not report technical problems with tools and technology since these issues are dealt with by a teaching assistant and thus not visible to the teachers. In this particular case, cognitive overload of students might be caused by using technology that is mandated by the stakeholders. However, in our experience, the challenges mostly stem from the social and organisational issues of involving external stakeholders. However, issues of these kind are only rarely reported in literature, e.g., by~\cite{Boehm1998} and	~\cite{kornecki1997strengthening}. 
These problems have not been picked up by us when designing the model and guide and they have not become evident when filling them in, either. Only when discussing the relationship to related work, we discovered this as a crucial aspect. A future iteration of the guide could thus include questions specifically targeting the technology aspect.

While the guide may appear to focus on static issues, it also covered some temporal or dynamic issues like to which course moments certain challenges are associated, as well as information about milestones or deadlines.
As such, the model can be used as for overall course design in the same way as the ADDIE model (cf.~\sref{sec:related-work:systematic-approaches} and~\cite{bates2016teaching}). But it can also be used to plan, execute and evaluate a series of interventions --- or sprints --- of stakeholder involvement within the same course. In this way the model bridges traditional, waterfall course design with a more agile approach. When used to capture a whole course instance the model still helps in accepting change as it highlights risky aspects of the design where the proposed mitigation strategies can become handy to redesign the course activities on the fly. 

\paragraph{Model and Structure of the Data Collection Guide}
While scalability has been discussed as a specific issue with the structure of the model before, there are additional observations about the construction of the model and the guide in particular. It is important to mention that there is a learning curve involved with applying the model. It took some time for some users to understand the various questions and which kind of data is required to answer them. We therefore recommend to go through the guide thoroughly without answering the questions and identify important data sources. Once the teacher starts answering the questions, all required material should be available and the connections between the different parts of the guide should be clear. It is also helpful to have a printout of the conceptual model (cf.~\fref{fig:conceptual-model}) close by in order to track progress.


\section{Using the Model Constructively to Plan and Execute a Project Course}
\label{sec:constructive}

While the course descriptions in \sref{sec:application} and the results in \sref{sec:results} focus on the retrospective use of the model, we have also applied it in its constructive capacity to one project course. Our experience with this way of using the model is reported in this section.

\subsection{Course Design}

EDA397/DIT191 Agile Development Processes is a course given in the Master programs of both Chalmers Technical University (CTH) and University of Gothenburg (GU). 
It is also provided as a PhD course with different demands. 
In principle, it is organised in two tracks: lectures and project. 
Both tracks are organised in three sprints (3 weeks each) with clear sprint goals:
\begin{longitem}
\item \emph{Sprint 1:} Getting started
\item \emph{Sprint 2:} Getting work done
\item \emph{Sprint 3:} Theory and advanced concepts
\end{longitem}

\paragraph{Lecture Track}
The sprints in this course help to better align lectures and project.
In Sprint 1, up to three lectures per week offer an overview of agile methods, principles, and practices to enable groups applying those in their projects. 
In Sprint 2, the number of lectures is significantly reduced to allow students to spend more time on the project. 
These lectures focus on cross-cutting concepts such as the agile spirit and comparing agile methods with plan-driven and lean software development. 
The final sprint introduces advanced concepts currently not incorporated in the project: distributed agile and large-scale agile, as well as agile architecting and system development are hard to make part of the groups' project work in the current setup. 

At the moment, guest lectures are scheduled in the second sprint (goal: match students' project experience with practitioners' views) and third sprint (goal: provide insights from practitioners on advanced concepts that are not present in the project).
Guest lectures were found to be especially useful to tie concepts from the project course to the real world and to complement theoretical knowledge with first hand experience. 

\paragraph{Project Track}
The project aims to give students a practical environment to explore and experiment with agile principles and practices. 
Students work in groups of eight or less students. 
They interact with the teaching staff during an initial project conception meeting, during frequent open Q/A sessions, during sprint acceptance tests, and sprint reflection. 
The latter two are strongly aligned with assessment criteria (e.g., in reflection we check if agile practices have been used in sufficient breadth and depth) and occur at the end of each sprint. 

At the moment, teaching staff act as customers (in the past three instances for an Android app that should support agile developers) and agile coaches. 
This could be a role to distribute to external stakeholders.
We used the model presented in \sref{sec:model} to analyse the potential benefits and challenges of such a change.
 
\subsection{Observations of Applying the Model}
We applied the sixth iteration of the model to the planning of the Spring 2017 instance, in order to plan involvement of external stakeholders both in the project and lecture track.
While there was some repetition and a large number of questions, the overall experience was good.
Many beliefs and tacit assumptions could be externalised. 
{Examples of these include motivation and effort of external stakeholders, as well as potential benefit for students. 
As a Master level course, we expect our students to be able to critically reflect on what they learn. 
While there is value to get insights on how things are done in practice, the learning goals of this course go beyond that.}

For defining external guest lectures from industry, 
{we would recommend a lightweight variant of the model}, since many aspects of the model did not seem to apply. 
With respect to the involvement of external stakeholders in the project track, however, it offered good opportunities 
{for} reflection. 

\subsection{Results of Applying the Model}

Applying the model impacted the course design in a very positive way and, despite our concerns about repetition and waste, we found the time well invested.
In particular, we can highlight the following three take-aways.

\paragraph{Common language} As a somewhat unexpected result, the model gave us a common language to discuss the use of external stakeholders with respect to our teaching goals.
This enabled us to engage with other teachers in a deep exchange of experience. 
We also found it very valuable to read through the earlier applications of the model (cf.~\sref{sec:application}), in order to compare our expectations with past experience.
We experienced this facilitation of discussion among the teaching staff for this course, but also with responsible teachers of other courses, very valuable. 

\paragraph{Better alignment of guest lectures with course goals} The application of the model forced us to make the expectations we have towards guest lectures explicit. 
This in turn leads to a better setup, where for example the course plan for the current instance indicated how guest lectures will relate to the project and will be included in the exam.

\paragraph{Deferred involving external stakeholders in the project} 
With respect to the project part of this course, we thought that external stakeholders could either interact with the students as customers for the agile teams or as agile coaches. 
The result of applying the model to this part of the course was to not involve external stakeholders due to the risk of skewing the student focus towards product instead of process. 
While students probably do expect external stakeholders to give guest lectures, they will not expect external stakeholders in the project. 
In past instances of the course, students overall appreciate the course, but had one major critique about the project focusing too much on producing a working software product and not enough about applying proper agile practices.
Since an external stakeholder would increase the product focus, our students could be distracted from the agile learning outcomes and prioritise the product quality over learning new agile practices. 

Thus, we feel that there is significant risk to the course as well as to the goals of the potential external stakeholders, as opposed to relatively small potential gains. 
Adding an external stakeholder in parallel to addressing this issue would multiply the risks. 
Students of this course are half way through their Master studies and will be in industry soon (or even return after some time in industry). 
Thus, we need to prioritise the course goal of learning and reflecting on agile methods over better insights into real world development constraints at this point of time.
 
In 2016, we added a reflection phase to complement the acceptance test after each sprint to better emphasise the importance of agile practices in the project. 
In 2017, we in addition require a written report on each of the required practices as input to the sprint acceptance test and reflection. 
We are confident that this will help to ensure our main learning goals and give us the freedom to start including external stakeholders for individual projects when time is ripe.

\section{Conclusion and Future Work}
\label{sec:conclusion}

While including external stakeholders in capstone projects and other project courses can be beneficial for many reasons, it comes with multiple challenges, such as misalignment between stakeholders and students or disappointed expectations.
To circumvent these challenges and increase the benefit of stakeholder involvement, we presented in this paper a model to analyse stakeholder involvement in project courses both retrospectively, i.e., in past courses, and constructively, i.e., for future course instances.
We described in detail the process of creating the model in six iterations in a ``guild of teachers'', i.e., through discussions among teachers.

Furthermore, we evaluated the model by applying it retrospectively to eight university courses and constructively to one additional course.
Our findings show that the model is suitable to retrospectively analyse the involvement of external stakeholders in capstone projects and project courses. The analysis of an additional course indicates that it is suitable to constructively plan the involvement of external stakeholders in capstone projects and project courses.
Thus, we present a model that can be used to model the involvement of external stakeholders in course design, answering research question \emph{RQ1}, and discuss strengths and weaknesses of the model, answering research question \emph{RQ3}.

We observe that many challenges in external stakeholder involvement can be summarised under common themes and occur in multiple of the analysed courses. For example, we observe that misalignment between students and external stakeholders is common, e.g., due to differing terminology used by teachers and external stakeholders. For each of these themes, we propose mitigation measures that have either been applied successfully, or have been planned for future course instances. However, it is important to note that their success depends crucially on context, forces, and constraints such as the course setup, the university environment, and scheduling.
While many of the observed challenges were known to the teachers before, the identified risk themes help to better understand causes for the risks and reason about appropriate mitigation strategies in a structured way (answering research question \emph{RQ2}). 
Identifying the risk themes was possible due to applying the systematic way of analysing the stakeholder involvement to nine courses.



In future work, we plan to investigate whether a more high-level and adaptive iteration of the model can help to mitigate identified weaknesses, such as scalability problems (see the discussion in \sref{sec:results:lessons}).
Another weakness we experienced is the dominance of the teacher lens due to the fact that it is more difficult to collect enough data from other lenses, e.g., from students and external stakeholders. 
To address this in the future, we plan to investigate how the model can be used to collect more feedback from students, stakeholders, and the university.
It will also be interesting to evaluate the model in a wider range of courses and contexts, e.g., considering more courses outside Sweden and courses in other computer science disciplines than software engineering. While we believe that the model is applicable to computing education in general, we do not currently have evidence to support this claim.
Furthermore, we plan a long term investigation of the model, applying it to our courses across multiple course instances, both constructively before and retrospectively after each instance. 
Finally, it is worth considering the impact of moving stakeholders between courses in future work.

\bibliographystyle{ACM-Reference-Format-Journals}
\bibliography{external-stakeholders}


\elecappendix

\medskip

\section{Full Course Descriptions}
\label{sec:appendix:coursedescriptions}

\subsection{DAT255/DIT543: Software Engineering Project}
\label{sec:appendix:dat255}

\subsubsection{Course Design}
The Software Engineering Project is a second or third year course taken by students from the Computer Engineering, IT, and Industrial Economy (specialisation in IT) bachelor programs at Chalmers Technical University (CTH) and the Computer Science program at the University of Gothenburg (GU). The course is given twice a year, during the spring and autumn terms to around 60 and 140 students respectively. While the student population is relatively homogeneous in the spring term with students mostly from the Industrial Economy program, the larger group in the autumn term is from a variety of programs with a more technical background. The course runs half-time for a total of 10 weeks in each instance, equivalent to 7.5 ECTS. In the spring, it runs in parallel with the students bachelor thesis project, giving our course lower priority. To make matters worse, public holidays during Easter and Pentecost need to be taken into account as well as a re-examination week and the university's centrally mandated deadlines and presentation dates for theses. There are less restrictions during autumn, but university deadlines and other courses still need to be considered.

The course is centred around a project where students develop a software product in teams of five or six students. Learning objectives are, however, not focused on assessing programming skills, but rather emphasise process-related issues, including understanding stakeholder requirements and selecting the right software engineering methods for the task at hand. This emphasis is also obvious in the teaching moments: there is a strong focus on process aspects, in particular on skills to analyse stakeholder requirements and plan the work accordingly. Scrum, with its iterative-incremental development lifecycle that allows regular insight into student progress, is chosen as a software development process.

External stakeholders are used in different functions in the project course, depending on their availability and the concrete project the students work on. In most cases, external stakeholders act as product or case owners. That means that they set the requirements for the product the students are asked to create and are also the ones to review the final results. External stakeholders are also used as mentors, in particular for helping students with the application of the software development process. In addition, guest lectures that connect the teaching material to practice are a part of the course. Finally, external stakeholders have been involved in other capacities, e.g., as a jury that evaluated the students' products in the context of an innovation competition \cite{EVS29,AMCIS16}.

\subsubsection{Observations of Applying the Model}

Two iterations of the model have been applied to the course. Every time the model was used, it was employed in a retrospective fashion, analysing all previous course instances together. The first time, model iteration 3 was used. While the model proved to be very helpful in identifying the different kinds of stakeholders that are involved in the course, it was difficult to identify their respective aims and expectations. It was also observed that the perspectives in the analysis part were not always clear. That was the reason to describe most aspects purely from a teacher perspective, rather than trying to understand and incorporate the perspectives of the students, the stakeholders, and the university as well. 

The second model that was employed was iteration 6  that included the guide. This proved to be a much more fruitful exercise since the questions in the guides helped in focusing the discussion and made it easier to consider the different perspectives. In the beginning, the guide was slightly confusing since the questions are formulated mostly for use in the constructive approach. However, after getting used to this, it was easy to answer the questions retrospectively. Some questions, such as the ones on the ``mode of interaction'' are more helpful in constructive use when assessing different alternatives, but these could easily be skipped. It was helpful to know the guide in advance, because that allowed jumping ahead and putting information in the right place without answering similar questions multiple times. Some of the questions do initially feel redundant, but since they are asked in different parts of the questionnaire and often with different lenses in mind, duplicate answers can be avoided if the guide is analysed and understood beforehand.

\subsubsection{Results of Applying the Model}

Applying the model helped understanding some shortcomings of the current course design that were unrelated to the involvement of external stakeholders as well as identifying challenges with the way external stakeholders are integrated in the course at the moment. An example for the former is that there has so far been no alignment of the learning objectives of the project course with other courses in the different programs. While this was implicit knowledge by the teachers, it has never been formulated explicitly. Work on the model has thus triggered a number of discussions and efforts to improve how the course is embedded in the different programs it caters to.

When focusing on external stakeholder interactions, the model helped understand the different aims from different perspectives, but it also made it clear that there was currently no means to gauge whether these aims have actually been reached. Thus, a crucial step in the process is currently missing. This is a clear area of improvement. It also became clear that the reason the external stakeholder involvement worked reasonably well in the project course was that there was a clear mandate for all external stakeholders to represent their company and resources devoted to the course. However, this requirement was specifically enforced by the teachers due to poor experiences based on a lack of management commitment in the past. 

The model revealed that external stakeholders were finding themselves in a new role they were insufficiently prepared for as a potential problem. Even though external stakeholders, e.g., the process mentors mentioned above, were briefed by the teachers, they often deviated from the briefing. In one extreme case, the mentor told students to not use Scrum, the software development process applied in the project. In general, a mismatch between the aims of the external stakeholders and the teachers' aims was observed. The main (perceived) aim of the external stakeholders was to represent their company as a desirable place to work and potentially recruit promising students. The teachers aim was to include practical aspects in the teaching and to have mentors help in teaching the process. Unfortunately, the external stakeholders were often too busy talking about their company to support these aims. Student aims and external stakeholder aims generally matched since the students could network with prospective future employers.

The main strength of the model applied to this course was the support it provided in identifying risks.  It was possible to identify communication issues that caused objectives to not be clearly formulated, lack of commitment that caused external stakeholders to prematurely drop out of the course, confusion for the students due to contradictory statements, among others. Using the different perspectives (teacher, stakeholder, student, etc.) helped in analysing these issues systematically. The translation into an action plan was more difficult and the defined mitigation strategies remained somewhat abstract. In particular, the lack of a measurement strategy to determine the effectiveness of mitigation strategies, already mentioned above, became apparent again. 

Since concrete data about the experience of the students and the external stakeholders was sparse, the usefulness of the observation part of the model was also limited. Only snippets from the course evaluation, the teachers' subjective experience and the limited communication with the external stakeholders could be used. One notable challenge that was discovered was that the mentors felt that they did not reach their original objective (recruiting). However, mentors were willing to come back in other roles, e.g., as case owners. When discussing the meaning of these observations, the model triggered a number of discussions. The teachers agreed that the experience with external stakeholders is overall positive. The integration with other courses in the study program needs to be improved.

\subsection{PA243: Master Thesis in Software Engineering}
\label{sec:appendix:PA243}

The Master Thesis in Software Engineering (PA243) is a capstone course taken by students at the Software Engineering program at Blekinge Institute of Technology in Karlskrona, Sweden. The course is given once every year, in the winter and spring of the final semester of the program. The number of students in the course have varied depending on the admittance numbers for the program but for the three years in focus for this retrospectve analysis (2008 to 2010), it was around 40-60 student per year. The student population is very heterogeneous with three main sub-groups: students from Sweden, from Eastern and Southern Europe, and a large group of students from Asia, primarily India, China and Pakistan. The course runs nominally for five to six months with a start in January and presentation of the thesis in early to mid June. However, in practice many students take seven to nine months to finish their thesis. Ambitious students tend to start early (October to November in the year before the formal start) and finish `on time' while many students run over and present their thesis in September or October. The course is the last of the program and a full-time course and thus have no restrictions from other courses. The course is restricted by the academic year though since no courses can be examined and formally finished in the summer, roughly from mid June to mid August.

Except for a few early lectures to support the students the course is almost entirely made up of individual projects that students do individually or in pairs of two students. The master program where the course is the final part prides itself with a close connection to the surrounding industry and the whole university has `Applied Information Technology' as one of its main foci. There is thus an implicit expectation from both students and teachers that the thesis projects will be done in collaboration with one or more companies. Since the research group for Software Engineering at the university is an internationally well-known one there is a small group of students that want to continue with a PhD after their thesis project; they sometimes prefer a more research-focused project done in an academic lab. However more then 85\% of students aims to do their thesis project at a company and most of them also manages to find one or more companies to host their project. Thus there is a very large variation in the type of project that students conduct. Some are more technical and aims to develop and try a new technology or tool within the industrial context, other focus more on the development process and/or help the company evaluate a new method within their development projects. Yet a third type is close to research and surveys or observes developers at the company in order to support improvements in the company's ways of working.

Overall, the course learning objectives is for the students to show their maturity and breadth of knowledge in Software Engineering and that they can build on and use this knowledge for advanced research and development projects in practice. This allows for the large variation that is seen between the student projects and also means that there are relatively few fixed requirements on the actual process and timing of the students work in the course. However, with support from the Swedish government the course went through extensive development in the years from 2007 to 2009. At the end of the analysed time period the course thus had fixed dates early on for when an extensive project proposal had to be approved, as well as a fixed process and deadlines for when the actual thesis had to be submitted and evaluated before the final thesis presentation. Four detailed rubrics to support the evaluation of the proposal, the thesis, the presentation and also the thesis process itself was also developed to help create more clarity for students and a more common basis for judging their work, by teachers. 

The two main types of external stakeholders for projects in the course are software engineers or managers in a company or a senior researcher in Software Engineering. Since a majority of projects are done in industry we focus on the former, below. Since the projects are also very different for different students there is also a lot of variation in the number and type of external stakeholders involved. Some students are `embedded' in the company, sits there througout their project and have daily contact with one or a few company employees. Other students visit one or a small set of companies more rarely while having mainly telephone or video conference interaction in between visits. The level of involvement of the company and the external stakeholders also vary widely between projects. However, all (non-research) thesis projects have a main supervisor in the company and an academic supervisor at the university and it is required that they have at least one early meeting together with the students to clarify expectations on the project.

\subsubsection{Observations of Applying the Model}
We applied the initial and the last iterations of the model to the course. Both models helped us better understand and analyse how the external stakeholders were used and involved in the course. They also helped in understanding how the stakeholder involvement evolved over the three years and suggested reasons for why some of the changes introduced in the course worked, or not. While the final iteration of the model had more detailed questions and thus supported a more detailed analysis of certain aspects of stakeholder involvement the large variation of thesis projects meant that it was perceived as more `costly' to apply this iteration to the many different stakeholder types. The initial, and more `loosely' structured iteration of the model allowed for more creativity in the analysis. The analyst could also adapt the level of detail in the descriptions to areas that was deemed more relevant for the course. However, the result of this analysis was harder to compare to results from other courses and requires more background knowledge from the analyst. It is not clear that the initial model would have been enough to support an analysis if the project itself had not been developed and improved in the government-supported evolution project. Basically, the development project meant that the involved teachers had more time to put into the course as a whole as well as write reports on its evolution. There is thus a trade-off between a detailed and more structure model, which ensures a more comparable and detailed results, and a less detailed and unstructured model, which allows for more analysis freedom but also requires more detailed knowledge. To cater for such different analysis situations the model might need to come in different iterations one being a less detailed iteration of the `full' one. It should also come with instructions to allow teachers to adapt its use, and which areas to include, based on analysis goals and context. Obviously, there is a risk that we lost some detailed information about the course instances since we performed a retrospective analysis several years later; a more natural use of the model is right after a finished course instance.

\subsubsection{Results of Applying the Model}
The main benefit of the model was that it helped analyse the goals of different types of external stakeholders involved in the course. Since the thesis course has so many different types of stakeholders involved and many of them have key roles to play in a successful project the more detailed analysis prompted by the model helped creating a deeper understanding of which changes to the course worked or not. In a previous research study~\cite{Host2010SEThesisSupport} we focused on better understanding the problems and types of support needed of students and supervisors involved in the course. The model presented here helped clarify what are the goals of other stakeholders involved such as academic supervisors and the teacher responsible for the course, as well as the program and university.

The reflections that is prompted by the model also suggested that student and stakeholder feedback can be routinely collected during as well as at the end of thesis courses. Since the external stakeholder interaction in a thesis course does not happen in a central location or at a specific time web-based solutions can be used to survey students on for example which type of interaction with stakeholders they have and at what frequency. While the existing course development had focused on the evaluation of course moments and to ensure that basic support information and recommendation is communicated to students and stakeholders the `loop' that is explicit in the model presented here points to continuous feedback and data collection also during the course.

The retrospective analysis also helped uncover the fact that a repeated problem is that external stakeholders in thesis project that fails tends to blame the students, while students tend to blame the company. While not very unexpected in hindsight this has not been in focus in earlier analyses that was not clearly based on a model and based on analysing several years of one and the same course. This will affect future courses by putting more focus on describing different `modes of failure' from the different involved perspectives. Potentially this can be discussed in an early lecture of the course as well as communicted from academic supervisors to industrial supervisors in initial meetings. By creating a shared and common `language' for talking about failed projects it seems likely the involved parties are more likely to be able to avoid similar problems.

 \subsection{TDA593/DIT945: Model-Driven Software Development}
\subsubsection{Course Design}
Model-Driven Software Development is a Bachelor-level course that covers the basics of software modelling, especially the Unified Modeling Language (UML).
It is part of several programmes at CTH and GU, including the option for master students from the Software Engineering programme of CTH and GU to take the course as an elective.
The 8-week course consists of about 10-14 lectures and a project spanning the entire course period.
The project is organized according to the lecture content with weekly assignments that build on top of each other and weekly supervision.
The lectures in this course cover different UML diagram types, including their syntax and semantics.
They include practical examples and theoretical discussions on the use of models, their abstraction level, and trade-offs between different example models.
The learning objectives are (on a high level) that students learn to create UML diagrams for a given purpose and understand diagrams created by others.
Furthermore, they should be able to reflect on topics such as the choice of the diagram type, the abstraction level of models, and the trade-off between different levels of abstraction.
Towards the end of the course, there are typically two guest lectures by industrial practitioners on the use of models in industry.
The purpose of the guest lectures is to convey to the students the use of models in an industrial setting and the relevance of the topic.
This is currently the only external stakeholder involvement in the course.

\subsubsection{Observations of Applying the Model}
The application of the model was straightforward as there was a clear structure with questions for each part of the model.
While this made the model application easy, it was also time consuming.
Furthermore, it sometimes felt as if questions had been answered before.
Typically, this was due to misunderstandings in the wording of the question, something that we expect will change once the person applying the model is familiar with it and has used the model multiple times.

Prior to coding the application of the model, the outcome felt a bit too spread out.
However, this might be related to its retrospective use. 
When actually targeting course improvement, it should prove helpful.

\subsubsection{Results of Applying the Model}
The model application helped us to understand in more depth the different aims of stakeholder involvement.
From teacher side, we mainly want the students to see real-world relevance of the course topics.
Therefore, we consider guest lectures from industry an easy and low-cost way to expose students to industrial examples for the use of software modelling.
While the guest lecturers can be seen to share this aim, we also observed that in some cases they are simply giving the course responsible a favour when giving the guest lecture.
Hence, their motivation to give a good and educational lecture might not be as high as it could or should be.
Finally, the students are motivated to attend guest lectures as they consider it beneficial for their career to understand real-world problems, but also for networking reasons.
That is, the aims are not always aligned.
While this is not a problem itself, certain aims can be seen to cause friction, e.g., the teacher's aim to provide students with a real-world view and the stakeholder's aim to do the course responsible a favour.

Prior to applying the model, we knew that while there is positive feedback from the student side, the industrial guest lectures are sometimes perceived too abstract, unprepared, or unrelated to the remainder of the course.
The model application and the discussions with the other authors of the model helped to understand this in more depth.
Indeed, we learned that there is a misalignment between course contents and the industry lectures, as the stakeholders are not well-prepared in many cases.
Additionally, they are not in detail aware of the structure and contents of the course.
This means that their guest lecture is often detached from the course context.
When this misalignment occurs, many students have difficulties to understand the connections and thus risk to not reach the intended learning outcome, i.e., to understand the real-world application and benefits of software modelling.

While the misalignment of guest lectures and the remainder of the course seem to be at the heart of the problem, there are several risks that directly follow from this.
First, students might lose motivation to attend future guest lectures or other lectures in the course.
This can then jeopardize their performance regarding other learning outcomes of the course and their performance in terms of grading.
Secondly, not understanding the industrial relevance of the topic might lower their overall impression of the usefulness and relevance of the taught topic.
From course evaluation surveys in the course, we see that this risk indeed manifests regularly.
Thirdly, the course evaluation might be impacted in a negative way, effectively leading to a lower evaluation of the teacher because of his/her efforts to teach the students the real-world impact of the topic.
Finally, all of these risks could lead to disappointment on the teacher side and a lower motivation to integrate external stakeholders or, in the worst case, to teach the topic or the particular course.

To mitigate the named risks, several mitigation strategies were explored using the model.
First and most importantly, stakeholders need to be prepared.
This means that they need to be aware of the students' backgrounds, e.g., their knowledge of subject matter or of the stakeholder's domain; the course contents and the depth in which these are taught; and the expectations towards the guest lecture.
The expectations explicitly also include the amount of effort the stakeholder should put into preparation of the guest lecture.
While this might not always cause the stakeholders to actually invest the necessary effort, it at least gives them an indication and might lower the risk of the stakeholder to just give a lecture in order to do the course responsible a favour.
This first mitigation strategy mainly covers the teacher-stakeholder relationship.
As a second strategy, the teacher-student relationship needs to be targeted.
Even if the stakeholder is well-prepared and gives an excellent lecture, it might be difficult for the students to understand the relation to the remainder of the course.
Therefore, a follow-up lecture given by the course responsible or a discussion at the end of the actual guest lecture is needed to relate the discussed concepts to the course.
The course responsible needs to directly align the stakeholders examples, terms and other concepts that might be unclear to students to the terminology used in the course.
This helps students to understand the connection points and the gap between industrial topics and the taught material.
Finally, as a more pessimistic strategy, industrial lectures could be skipped altogether.
While this removes the potential learning outcome of understanding industrial relevance of a topic, it also mitigates the entire range of risks associated with the guest lectures.

\subsection{DIT029: Software Architecture for Distributed Systems}
\subsubsection{Course Design} The project course Software Architecture for Distributed Systems is worth 15 credits and is given to second year Bachelor software engineering students over the course of a semester at the University of Gothenburg. About 70 students have attended the course. Students work in teams of 6-7 people to develop distributed systems. There were 10 student groups of 6-7 members. A student supervisor has been appointed to each team. The teams interact with an external company based in Gothenburg, the Techno Creatives AB.

All developed systems must communicate through a shared broker (i.e., MQTT). The choice of a project idea is free, but as a requirement each project needs to have at least one component that communicates over the network through the shared broker. The intention is that the created systems, while being self-contained, also communicate with each other using the protocol.

As any useful interaction of systems must involve a protocol, any student group that wants to publish messages on the shared broker must follow standards, which are published in the form of RFCs (Request for Comments), modelled after the RFCs that specify the traffic on the Internet. Each group needs to propose at least one RFC and have it approved, and also implement at least one own standard RFC in addition to one submitted by another team.

The objectives of the course can be summarized along three main goals:
\begin{longitem}
\item Students are expected to apply software architecture principles and practices to an industrial distributed systems case.
\item Students are expected to work in agile teams that involve an external company which provides regular feedback.
\item Student teams are expected to identify opportunities for cooperation with other teams, while working on objectives, which are specific to their projects. The cooperation might involve using services provided by another team, providing a completely new service together, or using a common solution to satisfy a recurring technical need. The whole class is expected to work in an ecosystems setting.
\end{longitem}

The idea of using MQTT and RFCs for project development came from our industrial partner. RFCs are approved by the company on a biweekly basis. An RFC that gets standardized should contribute to a common goal of interconnectivity between clients and specify a communication protocol or an aspect of it. Meetings between teacher and students were held biweekly at the university premises. Interaction and communication with the company have been organized remotely due to limited of face to face availability of the company. To facilitate the work of students, the company offered the technological platform to be used. It was agreed that the company evaluates the proposed RFCs by students but does not assess student learning outcomes. A final demo event has been organized at the company premises.

\subsubsection{Observations of Applying the Model}
We applied V6 of the model retrospectively to one instance of the course. While, applying the model has been straightforward some sections/questions have been found repetitive and overlapping. The reason could be that most issues raised are discussed from the perspective of different stakeholders in different sections of the guide. When addressing a specific question, say observed problems from the perspective of students, it is often difficult to separate between what we teachers think and what students have actually reported. Ideally, the guide should have been filled by different course stakeholders: teachers, teaching assistants, students, and external company.

Another interesting observation of applying the model has been to iteratively revisit previously answered questions every time a new issue is addressed. In many cases an answer has to be edited by removing, adding or modifying some part. This is a further illustration that the guide addresses cross-cutting concerns that are difficult to decompose and isolate.

Applying the guide offered a structured way of analysing the course experience, documenting the major issues, and identifying remedial actions to manage the risks. The guide may sound rather static in nature. However, when answering the questions often temporal and dynamic aspects of the course structure, such as milestones and deadlines, were also captured. For instance, risks identified are often associated with specific course moments held in the beginning of the course.

\subsubsection{Results of Applying the Model}
A number of risks have been identified in the Software Architecture for Distributed Systems course. At times students had to wait for a long time to get feedback from the external company on their proposed solutions. Such delay however is thought to be reasonable for the company. According to them this is how ecosystems work in practice after all, which is in itself an educational experience. Their availability as keystone is rather limited and it is natural to rely on the contribution of all actors. Accordingly, the company expected student groups to give feedback on each other's works and not just wait for company input. This resulted in a misalignment between student and stakeholder expectations. As mitigation strategy, one could consider to supervise student groups more closely to help them understand the expectations of the external company. Such strategy would promote higher levels of transparency and coordination between company and students. This however might also mean that extra teaching resources are needed to steer the collaboration.

Furthermore, the company has been supplying students with extra course material that did not directly relate to the course content but to the infrastructure offered by the company, which is not formally part of the course. It has been challenging for some student to learn and use the new resources when developing their solutions. The architectural style to be used as part of the developed solutions was introduced by the company in the beginning of the course but its theoretical foundation was discussed by the teacher later in the course. The company wanted to test how students perceive such an architectural style in early development phases. This has resulted in a misalignment between course content and industry lectures. In addition, the company mentioned to students that they are hiring (smart students) which was perceived as extra pressure. Those risks could have been addressed by minimizing the additional objectives introduced by the company.  In practice, this might mean that the company should have exercised less influence on the course. However, low influence level might also discourage the company to continue the collaboration.

Finally, many students thought that they are strongly dependent on company infrastructure and feedback, which would negatively affect their learning and grades. Such feedback has been often channelled through teaching assistants which was in some cases a source of confusion and misunderstanding between all parties. Even when encouraged to make direct contact, many students felt unease to talk directly to the company. The online communication platform has rarely been used by students, which meant that more meetings should have been scheduled with the company. However, this was not an option due to limited availability and resources. In this regard, the company had to cancel the halfway demo session since they had to attend an important meeting in another country. As a mitigation strategy, more regular workshops with the company should have been planned during the course planning phase. The plan should take into consideration possible cancellation. However, one should avoid too many meetings otherwise students would experience even more involvement and pressure from the company. On the other hand, in case meetings are cancelled during the running of course, students might complain about too many changes in the course plan.

When planning the first instance of the course, we thought that involving an external company in the course is all positive. However, it turned out that there were many unforeseen problems in addition to the expected advantages. For instance, we thought that the opportunity to be hired by the company is a motivational factor for students, but it turned out to be a source of pressure for some students. We also overlooked the fact that some students might feel being experimented by the company. In addition, feedback from the company was often channelled through teaching assistants which was in some cases a source of confusion and misunderstanding between all parties. Therefore, the purpose of the applying the model has been to plan the next instance of the course taking into consideration the identified risks and mitigation strategies.

We are working on an improved course implementation that better aligns the goals and expectations of the different course stakeholders. The aims and expectations identified from the different lenses will help us in such planning.

Applying the model to other courses taught by other colleagues in different programmes and countries helped identifying concerns and solutions applicable to the course. For instance, risks such as “Students are too strongly dependent on stakeholders” and “Working on a new open problem, with new tools and processes can create cognitive overload among students” identified in other courses were found very relevant to our course too. The same thing applies to solutions. For instance, we will implement mitigation strategy “Create support material for Students and external stakeholder” as reported in other courses.

Another result of applying the model is to communicate the experience of the first instance to the different stakeholders highlighting both the success stories and observed problems. We believe that this will contribute to bridging the gap between the expected course learning objectives and the actual learning outcomes experienced by students.

\subsection{DAT265/DIT599: Software Evolution Project}
\subsubsection{Course Design} The course Software Evolution Project is a shared course between the University of Gothenburg and Chalmers University of Technology that is taken by students in the last year of their master’s studies of the software engineering programs. There are often single students from other programs in Computer Engineering and IT. The course is given once a year in fall and spans as a 15 point course both study periods, i.e., the whole semester. Students are expected to invest 20 hours a week on that course. Parallel to the course the students take per study period one selected course. Many students also work. During the semester there is a break between both study periods, during which students typically write exams for other courses.

The learning objective of the course is to train students in software evolution and maintenance tasks, which includes comprehension of legacy code that has not been written by the students, the identification of (quality) improvement potentials, the implementation of improvements to the system, and the implementation of a typical evolution scenario, such as a platform migration. To achieve that the students work on a real software product and are, especially at the start of the course, coached in a number of techniques that can support software comprehension, the identification of quality flaws, and refactoring.

To improve the learning outcome it is crucial to provide the students with access to a real system with real flaws. We address that by using open source systems. The two here discussed instances differ in the way the open source systems were selected. In the first instance students chose these systems themselves freely, while we restricted the choice to two systems in the second instance. In both years the contact with these real systems helped students get a taste of typical challenges with real systems. However, students themselves had own aims, e.g., making a real impact to these systems. As it turned out, this aim is more difficult to achieve.

\subsubsection{Observations of Applying the Model}
The model was applied to two instances on the course (2015 and 2016).
We did that for three different iterations of the model: 3, 5, and 6.
Applying the model in the final iteration took roughly 1 hour, only. However, there was of course some reuse from the answers to the previous iteration (applying the model iteration 5 took about 2 hours). The guide is now easy to apply (most unclear parts are resolved, double question removed). However, this impression might also be the result of experience, e.g., it probably gets easier to fill the guide for someone who has done it one or two times already.

An advantage of the model is the guide with relevant questions. First conflicts that I observed during the courses were popping-up during filling-out the guide. For example, the questions for students' side-activities made it directly clear that there is a potential that teams might have trouble to regularly meet. This is a problem that we actually observed during the course.

To sum up, the model is a real help to document and cover the aspects of the stakeholder interaction and course planning and helped to trace back some observed problems to risk factors in the course set-up. However, not all questions might be relevant for all courses. Thus, the user needs to be a bit selective when filling in the guide. Apart from that the models seems to be in a good position to allow for a comparison of courses. 

\subsubsection{Results of Applying the Model}
Applying the model helped to make several risks explicit. For example, we observed that there is a strong and inherent misalignment between the way open source communities welcome and accept newcomers in little steps and the goal to make a real contribution during a one semester course.  Furthermore, we could observe that students seem to need more technical support than the one they can get via the available documentations. Support from the OSS projects does seldom consist of deep feedback.

Furthermore, applying the model helped to reflect about how changes made between the two course instances actually helped to mitigate some of the risks. For example, adapting the assessment allowed us to reduce the risk that students target the wrong goals in order to increase their chances of getting their contributions accepted by the OSS project.

To conclude, since this is only a reflection form the teachers’ perspective, it is not entirely clear which of the remaining risks have the strongest impact on the students. However, we can now start to reason about the risks and address them step by step, which was not possible before using the model. For example, we will discuss several alternative strategies now that might help us to ensure that students can gain enough technical support if they need it. This might either be through training course supervisors in the open source systems ahead of the course start. An alternative could be to getting in touch with core developers of an open source system and agree on a collaboration there they would agree on actively supporting the students.

\subsection{CMPE 450: Software Engineering Course}
\subsubsection{Course Design} This was the only software engineering course offered to undergraduate computer engineering students in their fourth year at Bogazici University in Istanbul, Turkey until 2010. The goal of the course was to introduce students the phases of software development including feasibility study, analysis, specification, design, implementation, testing, documentation and maintenance. During this course, students also became familiar with software development tools, techniques, environments and methodologies as well as gaining knowledge about management issues. An integral part of the course was the involvement of students working in teams in order to develop a medium size software product. 

The model was applied to the instance of the CMPE 450 course that was taught in 2007.  The course was offered to approximately 80 students. Most of the students were from Computer Engineering undergraduate program (\~85\%) and only this instance of the course was also offered to students from Information Systems Engineering undergraduate program (\~15\%). As a software development project in this course, students developed a purchase management automation system to be used by the related units in Boğaziçi University (e.g., all academic departments, university’s presidency, budget unit, purchase unit, and all other support-units such as the personnel department, and etc.). In this course project, the external stakeholders were university’s employees including academic department secretaries, staff from the budget and purchase units, vice chancellor, staff from personnel department and student office. These stakeholders were the customers for purchase management automation system the students were supposed to design and develop. The rationale behind the selection of such a software development project was to teach students to design and develop software in real life settings.

\subsubsection{Observations of Applying the Model}
The final iteration of the model was applied to the course, retrospectively. Applying the model took roughly 5 hours. Since the instance of the course belongs to 2007, some of this 5-hour duration was spent to remember and retrieve from archives and some resources related information about the course. Some questions in the guide initially seemed redundant, however repetitions can be prevented as identical questions asked in different parts of the questionnaire need to be answered often with different lenses in mind. However, while applying the model I realized that my ``teacher's lens'' was more dominant while answering the questions. I tried to think myself in the shoes of students and external stakeholders, but how effective I was while applying those lenses is questionable. The guide was also helpful in highlighting the risks due to the course setup as well as in identifying actions to manage those risks. 

\subsubsection{Results of Applying the Model}
The strength of the model is that it helps you think thoroughly about the possible risks that might reveal due to the way you design and decide how to execute the course. I applied the model only retrospectively, but I believe that if I had the chance to apply the model while designing this course it would help me foresee the risks that I was able to identify later through retrospective analysis. It is also obvious that application of the model to a course in a retrospective manner yields outcomes that would help a course’s improvement of the future instances. 

Another strength of the model is that it helps the model-user show his/her peers risks involved in the current design of the course in a systematic and clear way. For instance, as teachers we already knew that the course content, which required working on a new problem with new tools and processes, caused significant amount of cognitive load on students. The teacher who was the main responsible of the course had already told the undergraduate program committee that the course content should be split into two separate courses. However, it took some time to convince the committee. Finally, in 2010 this course was replaced by two separate courses, which are “CMPE 352: Fundamentals of Software Engineering” and “CMPE 451: Project Development in Software Engineering”, respectively. During CMPE 352 course, students learn about software engineering concepts including software development methodologies and techniques as well as learning about tools for version management, issue management, software testing and code review. Requirements analysis and specification, and software design, implementation, testing, documentation and maintenance are among the other topics of this course.  CMPE 451 course gives the students the opportunity to put into practice what they have learned during the CMPE 352 course by working in groups in order to develop a medium size software product. It is likely that the committee could have been convinced more easily and much earlier if the model existed at that time and we were able to apply it. This would help us explain the problem in a more systematic and hence more explanatory and effective way to the committee.  In this way, the course could have been re-designed into two separate courses earlier than 2010. 

The main weakness of the model is that due to too many questions, it takes too long to complete the guide for model iteration 6.  I sometimes could not figure out the reason why some of the questions were asked. After filling in the guide I needed some time to grasp the big picture to extract output the model is designed to reveal (i.e., the risks, mitigation strategies, lessons learned regarding the design and execution of the course). However, this might be also due to the case that I took part in teaching this course in 2007, which is ten years ago. I also had to go through some archives and force my memory to remember the details about the course. This probably caused significant amount of cognitive load on me while applying the model. Hence, regarding the length of the model guide, maybe results of the applications using more recent courses whose information is much easier to remember should be taken into account.  On the other hand, I was able to apply the model retrospectively to a course that I co-taught a decade ago. This shows that one can apply the model to courses, which were taught quite a long time ago, which is another strength of the model.  Mankind has the problem of forgetting about the past and repeating the same old mistakes to the limits in his memory. Applicability of the model to such courses helps teachers remember success stories as well as the mistakes of the past that might otherwise fade away from our memory.  Such information can be quite useful in designing current courses.

Regarding lessons learned about the design of the course, the model showed me the actual risks of introducing real life software projects into courses to be completed by students.  It is crucial to prepare a plan for the maintenance of the resulting software and assess the feasibility of this plan under existing circumstances. One possible solution would be to include university’s IT staff that is expected to be responsible for the maintenance of the software, as additional external stakeholders. However, earlier engagement of the IT staff with the development of the software project during the course might not solve the problem. There might be external factors beyond teachers’ control such as the high workload of the IT staff that may cause reluctance in taking over the software project.  Therefore, it is crucial to have discussions with parties who are expected to take over the maintenance of the software in order to make a proper assessment about the feasibility of our plan.  Another risk regarding real life software projects is that end-users with no software engineering background may not come with realistic requirements to the students. As a risk mitigation strategy during the course, we monitored the requirements gathering and elicitation process of students with the customers and intervened when such unrealistic requirements confused the students.

\subsection{CMPE 451: Project Development in Software Engineering}
\subsubsection{Course Design} This course is offered to fourth year undergraduate computer engineering students in fall semester at Bogazici University, in Istanbul, Turkey. During this course, students are supposed to put into practice the concepts they learned during “CMPE 352: Fundamental of Software Engineering Course” (e.g., software development methodologies, tools, techniques and environments) by working in teams in order to develop medium scale software projects.

The model was applied retrospectively to the first instance of the CMPE 451 course that was taught in 2011.  The course was offered to approximately 80 students from Computer Engineering undergraduate program. During the course, project teams were given a selection of software projects that were made available for the SCORE contest that is organized as a part of ICSE (International Conference of Software Engineering) with the goal of promoting and fostering software engineering in universities worldwide.  

Each software project had a proponent who was responsible for communicating with the student project teams and providing them required information for the elicitation of the requirements. The main responsible teacher of the course was a software engineer who volunteered to co-teach the course and one of the co-authors was assisting him in teaching. Besides having more than 20 years of experience in industry and he was also an entrepreneur, who is the founder of a large-scale company that is specialized in development of Enterprise Resource Planning (ERP) software products.  This person was the main external stakeholder throughout the whole course. During the period the stakeholder co-taught the course, since he was taking some time off in between the process of selling his shares of the large-scale company and moving to USA to start up a new company, he had enough time to get involved with the course. As a result, he was an ideal candidate to teach this course. 

\subsubsection{Observations of Applying the Model}
Completing the guide for the sixth iteration of the model took around 4.5 hours. While answering the questions in the guide, there were some interruptions in between since the co-author had to context switch to complete some other tasks. There were too many questions in the guide and some questions seemed redundant, since answering some questions already resulted in having been answered some other questions. For instance, answering questions about students’ expectations from collaborating with the external stakeholder and whether any of the learning objectives of the course justifies inclusion of the stakeholder, the co-author who co-taught this course had mentioned that the stakeholder is a software engineer with 20 years of experience in industry and founder of a large scale company. In this way, the co-author had already answered question asking who the stakeholder is as well as the questions about the company the external stakeholder represents and his qualifications. However, the impression of the co-author is that this might be due to the nature of the course the model was applied in addition to the role of the external stakeholder in that course. Besides, repetitions while answering questions made easy for the co-author to resume after returning back to answering the questions in the guide.  

The co-author also had difficulties in answering the questions in the guide regarding the ``Action Plan'' for the next iteration of the course. The reason for this is that full-time involvement of an external stakeholder with such qualities was quite a special case and it is not likely to happen for the next iterations of the course. When the involvement of the external stakeholder changes, the co-author had the difficulty in figuring out what parts of the experience can be transferred to the next iterations of the course.

While completing the guide for the model, the co-author mainly focused on the co-teacher of the course as the external stakeholder, since he was the only stakeholder who remained throughout the whole course. However, while completing the model guide, the co-author realized that the SCORE contest proponents, who were responsible for providing the necessary input for the student project teams, were also external stakeholders. However, she could not include them into the guide as external stakeholders. One reason for this is that since the proponents did not respond to any questions of the students, they were excluded from the course and as a mitigation strategy at the mid of the course the teachers of the course started to play the roles of these proponents and fulfill the required responsibilities on behalf of them. Another reason for this is that the way the guide for the model is designed makes answering questions for two separate stakeholders who have different roles and responsibilities in the course quite difficult. An alternative might be going over the guide twice and answering the questions taking into account only one external stakeholder each time. The co-author could not do this as this required extra time she could not afford to invest later.

\subsubsection{Results of Applying the Model}
One main strength of the model is that applying the model also helps you identify different types of stakeholders and what kind of risks are introduced into the course due to the inclusion of each type of external stakeholders. Initially, teachers had decided that it was mandatory to take part in the SCORE project competition of ICSE and 10\% of the grading would come from Performance Evaluation of the SCORE contest client. However, the score contest clients were mostly unavailable, in other words they were almost inexistent from the perspective of the teachers and also students.  Many of them even did not answer the emails of the student project groups and were not involved with the progress of the projects. It is likely that they have information about the content of the resulting projects. As a result, SCORE clients turned up to be not good candidates for evaluating performance of project groups. Hence we had to exclude their opinions from the grading of the projects. This also caused some kind of frustration among the students. We teachers tried to replace the functionality of the SCORE clients and acted as the only clients, tried to guess what they might imply with the written requirements. We tried to give as much feedback as we can to each project group and this helped the students to get motivated. We also decided that result of the contest should not affect the grade of the students if the students could not be among the finalists of the contest.

The model also has the strength in helping you think beforehand who are the right candidates to be external stakeholders. Especially, the model helps identify the potential stakeholders whom teachers can get in touch and communicate when a problem arises. SCORE clients were not such stakeholders and hence they were not ideal for the course. It was possible for teachers to exclude them from being external stakeholders during the rest of the course, since teachers took over their responsibilities. However, in some cases it might be more difficult or impossible in the middle of the course to exclude external stakeholders without mitigating the consequences of such an action.  If the model is applied in advance while designing a course, it can help decide whether the teacher(s) will be able to intervene if the external stakeholder does not do what he is expected to do and also whether teachers’ intervention will be enough for the success of the course. Despite these strengths, the model needs some improvements. One weakness of the model is that while answering the questions in the guide, one should focus on a single stakeholder or stakeholders who have the same role/responsibilities in the course. For each stakeholder role/responsibility, the questions should be answered separately and sequentially. However, this requires quite an effort and time.

\subsection{CSC302: Engineering Large Software Systems}
\label{sec:appendix:CSC302}
\subsubsection{Course Design}
Engineering Large Software Systems was a third-year Bachelor Course offered at the University of Toronto (U of T) as part of their Bachelor's Degree in Computer Science.  The course was the most advanced purely undergraduate course in Software Engineering (there was one further SE course which was cross-listed, both graduate and undergraduate students could enrol).   As such, it was mainly taken by students who had a ``Specialist'' or Major in Software Engineering.    A full-time student load in the program was five courses stretching over one semester.   Many students were taking four other courses, and several students had part-time work.   The course had about 50 students enrolled, but less than 20 typically attended lectures.  

Students had previously taken coursed dedicated to Software Engineering and programming.  A second year course introduced them to basic Agile practices, while the precursor to this course ``Introduction to Software Engineering'', also a third-year course, was meant to introduce them to basic SE practices, with a focus on Agile.   The Large Software Systems course was intended to cover ``advanced'' topics, including Requirements Analysis, Model-Driven Design, Test-Driven Development, Project Planning, and Quality Analysis.   Learning objectives were not made explicit, as was the culture at U of T and many other North American Universities.  

Model iteration 3 (pre-guide) and 6 were applied to one instantiation of the course.   The course instantiation was  special in that the course was taught by a graduate student instead of one of the two regular instructors.   As such, she was adjusting the material and following the guidance provided by her seniors, in a low-time and high-pressure environment.  This set-up was not uncommon for course instruction at U of T.

The course project involved contribution to a real, open-source project, Violet UML, selected by the students via a vote.  Interaction with external stakeholders associated with that project was minimal, the instructor communicated with one of the developers via email, and response time was slow.  Minimized communication was in part because the code base was not very active and the number of developers involved was small.  

Still, it was a tradition of the course to involve external stakeholders via guest lectures.  The instructor did not have appropriate industrial contacts, so relied on the past lecturers to provide contacts and make the invitations.  Application of the model is focused on this involvement, particularly the participation of one invited guest lecture.  
\subsubsection{Observations of Applying the Model}

Applying the model iterations 3 and 6 took less than an hour, but the time for the latter iteration was likely reduced, as some of the information in the previous application could be reused.  For this particular example, with a focus on guest instructors, applying v6 did not bring many new discoveries when compared to v3.  Many of the questions in v6 were redundant or not applicable, the application process in this case was rather tedious.  Useful aspects, like the explicit focus on the variety of stakeholders and their goals, seemed to be more implicit and hidden in the v6 model.  There was some confusion, as the iteration 6 guide seems to be geared towards planning a new course, where as the intention was to apply the model retrospectively for analysis.   Still, the overall application of the various iterations of the model helped the instructor to see the experience in a new light.  
%
%

\subsubsection{Results of Applying the Model}
The model captured the disconnect between the industrial guest lecture content and the course content, an issue of which the instructor was already aware. For example, the course focused on many model-driven techniques, where as the industrial guest lecture from a prominent SE company told the class they never use models.  Because she relied on someone else to invite the lecturer, there was a minimum opportunity for coordination.  However, the process of developing and assessing the stakeholder model brought to light ideal mitigations to avoid this problem in the future:  either coordinating with the lecturer to ensure content is consistent with the course, prepping the students pre- and post-lecture to discuss the differences between the course content and the lecture content, and finally just picking a different guest lecturer.    The instructor had not thought of these mitigations at the time, but was also not so aware of the risk of content disconnect.    

Note that it's hard to tell if these mitigation strategies would be found via only the model alone, but instead come from the process of group design of the model, including exposure to the risks and mitigations of others.  

Applying the model brought to light some of the institutional issues with the involvement of external stakeholders,  both at the group and department level.  These were issues that had not occurred to the instructor previously.   The group/department did not have any explicit mandate or guidance on the use of external stakeholders, but there was an implicit understanding that inviting guest lectures from industry showed connection to industry, and exposes students to useful information.   This was particularly true in Software Engineering, and was less true across Computer Science as a whole.   However, there was no guidance or support for selecting or inviting such stakeholders.

Overall, application of the model revealed a lack of an overall governance for the program.   Although the course instance was the responsibility of the graduate-student instructor, and the past instructors had an implicit stake in maintaining course quality, as far as the instructor could tell, no one was in charge of the overall quality of the course or program. More meta-analysis of best practices across course instances was needed, and this was out of the scope responsibility for graduate students or sessional lectures. The model is helpful in bringing to light issues not only with the individual course but with the program as a whole.  
%
%

\section{Data Collection Guide for the Use of the Conceptual Model}
\label{sec:appendix:guideline}

\subsection*{Instructions}
\label{sec:appendix:guideline:instructions}

The data collection guide presented below is an accompaniment to the model shown in \fref{fig:conceptual-model}. It contains questions that allow the teacher to reflect on the way external stakeholders have been involved in a course (retrospective use) or will be involved in a course (constructive use). For the guide to be helpful, it will have to be tailored to the concrete course. Not all questions are applicable for all types of interaction. Especially in retrospective use, some questions might feel redundant.
We therefore encourage each teacher to carefully go through the guide before answering the questions and select those that seem most helpful. 

Before answering the questions, the teacher should ensure that a broad foundation of information is available. In particular, any data sources that might be helpful during answering should be accessible to the teacher. Such data sources might include course notes, course evaluations, interviews with external stakeholders or students, reflection reports, notes from discussions with other teachers, or documents detailing strategies of the study program or the university. If an answer cannot be given due to insufficient information, this should be clearly marked since it entails an additional risk.
If possible, going through the guide with student representatives and the external stakeholder is strongly encouraged.

All answers should be recorded in written form. After the questions have been answered, a pass through them in which common themes and recurring issues are identified is recommended. The risk themes presented in \sref{sec:results:risks-mitigations} can be used, but self-defined themes are also possible. 

One pass through the guide will take about two to three hours, depending on the number of potential stakeholders and accessibility of the information.

\subsection*{Lenses}
\label{sec:appendix:guideline:lenses}

We use six different lenses, or perspectives, when planning, executing
and evaluating external collaborations:

\begin{enumerate}
\def\labelenumi{\arabic{enumi}.}
\item
  The autobiographical lens which refers to your own perspective
\item
  The student lens
\item
  The peer, or other teacher, lens
\item
  The theoretical lens, also referred to as the literature lens
\item
  The external stakeholder lens
\item
  The program and university lens which represents the broader
  educational context of the course
\end{enumerate}

The first four lenses are the same as Brookfield's reflective lenses
(``Becoming a Critically Reflective Teacher'', 1995) while the fifth and
the sixth are our own addition. The lenses play different roles and come
with different weight throughout the collaboration.

\subsection*{Part A: Influences}\label{sec:appendix:guideline:part-a-influences}

Influences are factors that have an impact on the course and the
possible choices the teacher can make. Not all influences need to be
considered by the teacher.

\subsubsection*{1. Teaching Context}\label{sec:appendix:guideline:part-a-influences:teaching-context}

Teaching context contains information about the students that will take
the course, their prerequisites, and their previous exposure to external
stakeholders.

\begin{enumerate}
\def\labelenumi{\arabic{enumi}.}

\item
  How many students do you expect in the course instance?
\item
  Which relevant courses have the students attended before?
\item
  Which knowledge and skill levels can be expected from the students?
\item
  Do the students have any kind of experience in interacting with
  (other) external stakeholders?
\item
  Do the students have any expectations regarding collaborating with
  external stakeholders?
\item
  Is the student group heterogeneous (e.g., due to different programs)
  or homogeneous?
\item
  Which other courses are ongoing while the students participate in this
  course and how does this influence their workload?
\item
  Do the students work on the same or different assignments?
\item
  Will the students cooperate with the same or different external
  stakeholders?
\end{enumerate}

\subsubsection*{2. Forces}\label{sec:appendix:guideline:part-a-influences:forces}

Forces are external conditions that influence the course, are outside of
the control of the teacher but do not necessarily have to be met by the
teacher.

\begin{enumerate}
\def\labelenumi{\arabic{enumi}.}

\item
  Does the university have goals in place that require teachers to
  include external stakeholders? What do they mandate?
\item
  Is involvement of external stakeholders part of the evaluation of the
  course by university authorities? If so, how is involvement evaluated?
\item
  Does the university provide programs to get in touch with external
  stakeholders for teaching purposes?
\item
  Are there other courses with which the learning objectives need to be
  aligned?
\end{enumerate}

\subsubsection*{3. Constraints}\label{sec:appendix:guideline:part-a-influences:constraints}

Constraints are conditions that set strict limitations on the course.
Examples are the course schedule (e.g., when the exams need to be taken,
the number of session per week, etc.) or alignment with schedules of
other events (such as competitions).

\begin{enumerate}
\def\labelenumi{\arabic{enumi}.}

\item
  What is the duration of the course and when does it end?
\item
  How is the course going to be assessed?
\item
  How often is it possible to interact with the students without the
  external stakeholder?
\item
  How often is it possible to interact with the students with the
  external stakeholder present?
\item
  Are there external events that need to be considered in the planning
  of the course?
\item
  Are there deadlines imposed by the university that must be met by
  either the students or the teacher?
\end{enumerate}

\subsection*{Part B: Prepare stakeholder involvement}
\label{sec:appendix:guideline:part-b-prepare-stakeholder-involvement}

\subsubsection*{1. Pedagogical purpose, intended learning outcomes, and other aims}
\label{sec:appendix:guideline:part-b-prepare-stakeholder-involvement:pedagogical-purpose-intended-learning-outcomes-and-other-aims}

The involvement of external stakeholders in a course should be aligned
with the intended learning outcomes and with the pedagogical purposes of
the course and the program. There can also be other aims, e.g., from the
teachers, that can influence the decision to include external
stakeholders.

\begin{enumerate}
\def\labelenumi{\arabic{enumi}.}

\item
  Which, if any, learning objectives justify the inclusion of external
  stakeholders?
\item
  Beyond stated learning objectives, are there additional aims for
  including (or that influences the decision to include) external
  stakeholders?
  \begin{enumerate}
\item
  From the perspective of the course itself? For instance, things that are in
  line with overall aims of the course but not explicitly stated as
  learning objectives.
\item
  From the perspective of the students in the course? For instance, their
  employability.
\item
  From the teachers (personal) perspective?
\item
  From the (study) programs perspective?
\item
  From the university perspective? (for example in relation to vision,
  profile or policies)
\item
  From any other perspective (not covered above)?
  \end{enumerate}
\item
  How can the achievement of the aims be measured?
\end{enumerate}

\subsubsection*{2. Potential stakeholders}
\label{sec:appendix:guideline:part-b-prepare-stakeholder-involvement:potential-stakeholders}

There are often a number of possible stakeholders that can be included
in a course. These stakeholders should be enumerated and some facts
about them should be recorded.

For each potential stakeholder, answer the following questions:

\begin{enumerate}
\def\labelenumi{\arabic{enumi}.}

\item
  Who are the people that will act as external stakeholders?
\item
  Which company does the external stakeholder represent?
\item
  What is the availability of the potential stakeholder?
\item
  Which qualifications does the external stakeholder have?
\item
  Is there support by the management of the external stakeholder?
\item
  Does the external stakeholder have experience interacting with
  students?
\end{enumerate}

\subsubsection*{3. Goals and trade-offs for each stakeholder}
\label{sec:appendix:guideline:part-b-prepare-stakeholder-involvement:goals-and-trade-offs-for-each-stakeholder}

Once potential stakeholders are identified, their goals should be
assessed and an analysis of the alignment with the purposes from Section
B.1 should be conducted. This should also include an analysis of the
trade-offs that including a specific stakeholder would incur.

For each potential stakeholder, answer the following questions:

\begin{enumerate}
\def\labelenumi{\arabic{enumi}.}

\item
  What are the aims of the potential stakeholder?
\item
  Do the aims of the potential external stakeholder align with the aims
  as stated in Section B.1?
\item
  What are the potential advantages of including the external
  stakeholder in the course?
\item
  What are the potential disadvantages of including the external
  stakeholder in the course?
\item
  What is the cost of including the external stakeholder in the course?
\item
  Which restrictions does the external stakeholder impose on the course
  and/or the students (NDA, subject, domain)?
\end{enumerate}

\subsubsection*{4. Action alternatives}
\label{sec:appendix:guideline:part-b-prepare-stakeholder-involvement:action-alternatives}

The final step in preparing the involvement of external stakeholders is
to define action alternatives and evaluate them. These action
alternative should be aimed at identifying how different potential
stakeholders can be involved in order to leverage the potential
advantages and prevent the potential disadvantages. It is sometimes
possible to choose different alternatives at the same time.

For each potential stakeholder, answer the following questions:

\paragraph{1. Type of involvement}

The type of involvement defines what kind of ``services'' the external
stakeholder offers to the students. The stakeholder could, e.g., provide
resources such as meeting locations, food, or build servers, supply
professional expertise, domain knowledge, organisational circumstances,
act as a mentor, as the product or case owner, or be involved in
supervision.

\begin{enumerate}
\def\labelenumi{\arabic{enumi}.}

\item
  Which type of involvement can the stakeholder offer?
\item
  Which type of involvement is suitable to reach the aims?
\item
  How much time and effort does the type of involvement require from the
  stakeholder?
\end{enumerate}

\paragraph{2. Mode of interaction}

The mode of interaction defines how the students interact with the
external stakeholders. Face to face meetings are possible, but also
online forums, email, phone calls, and others.

\begin{enumerate}
\def\labelenumi{\arabic{enumi}.}

\item
  Which modes of interaction are available for communication with the
  stakeholder?
\item
  Which mode offers the greatest value to the students?
\item
  Which mode offers the greatest value to the external stakeholder?
\item
  How do the different modes of interaction align with the aims?
\end{enumerate}

\paragraph{3. Risk assessment}

Involving third parties in a course has inherent risks that can, if not
properly addressed, jeopardise the students' ability to reach their
learning objectives. It is therefore crucial to make the risks explicit
and define a mitigation plan beforehand. The severity and likelihood of
the risks can also be an important factor in deciding which potential
stakeholder should be involved in the course.

\begin{enumerate}
\def\labelenumi{\arabic{enumi}.}

\item
  What are risks that come with involving the potential stakeholder?
\item
  What is the impact on the students and the course if these risks
  manifest themselves?
\item
  What is the likelihood that these risks manifest themselves?
\item
  What are potential mitigation strategies for these risks?
\end{enumerate}

\subsection*{Part C: Action Plan}
\label{sec:appendix:guideline:part-c-action-plan}

Once all possibilities are defined, the final preparatory step is to
create a concrete action plan and discuss this plan with the selected
stakeholders.

\begin{enumerate}
\def\labelenumi{\arabic{enumi}.}

\item
  Who of the identified external stakeholders should be involved in the
  course?
\item
  Which type of involvement should each of the external stakeholders
  offer?
\item
  Which mode of interaction should each type of involvement for each
  stakeholder have?
\item
  What is the schedule for the different types of involvement?
\item
  What are the relevant risks in involving the selected stakeholder?
\item
  How can these risks be mitigated should they occur?
\item
  Which measurements can be made to observe if the involvement of the
  external stakeholders achieves the intended aims?
\end{enumerate}

\subsection*{Part D: Observation and Analysis}
\label{sec:appendix:guideline:part-d-observation-and-analysis}

Once the action plan is put into place and acted on, data should be
collected according to the identified measurements. This data should
then be analysed following a set of lenses, inspired by and extended
from Brookfield.

\subsubsection*{1. Observation}
\label{sec:appendix:guideline:part-d-observation-and-analysis:observation}

The observed data should be reported as neutral as possible without
appraising them, assigning value, or meaning to it. Data from student
evaluations, post-mortems with the external stakeholders and a teacher
diary can be valuable sources of information for this purpose.

\begin{enumerate}
\def\labelenumi{\arabic{enumi}.}

\item
  What could you observe w.r.t. the behaviour of the students?
\item
  What could you observe w.r.t. the satisfaction of the students?
\item
  What could you observe w.r.t. the behaviour of the stakeholders?
\item
  What could you observe w.r.t. your ability to teach?
\item
  Where the interactions between students and external stakeholders
  aligned with your aims?
\item
  What could you observe w.r.t. other courses in the program?
\end{enumerate}

\subsubsection*{2. Meaning}
\label{sec:appendix:guideline:part-d-observation-and-analysis:meaning}

By using the extended set of lenses, the next step is to derive meaning
from the observed effects. Through the lenses, the meaning becomes
subjective for the individual considered.

\begin{enumerate}
\def\labelenumi{\arabic{enumi}.}

\item
  Autobiographical: What are your personal feelings about the
  involvement of the external stakeholders? Have you reached your aims?
  Where you able to teach the way you wanted?
\item
  Student: Did the students gain value from the involvement of the
  stakeholders? What were the issues they had?
\item
  Stakeholder: Did the stakeholders gain value from their involvement?
  How did they perceive the interaction with the students? Would the
  stakeholders change anything?
\item
  Peer: How did other teachers perceive the involvement of stakeholders
  in the course? Was there a positive or a negative impact on other
  courses in the program?
\item
  Theoretical: How do the effects and the meaning you derived from it
  compare to reported accounts? Did others see the same effects?
\item
  University: Did the involvement of external stakeholders generate
  value for the university? Were university goals reached?
\end{enumerate}

\subsection*{Part E: Reflection}
\label{sec:appendix:guideline:part-e-reflection}

In this final part, the effects and the meaning are used to reflect on
the experience and to derive lessons learned as well as proactive
measures that can be applied in future course instances. The reflection
is guided by three questions:

\begin{enumerate}
\def\labelenumi{\arabic{enumi}.}

\item
  What was? E.g., the meanings of the collaboration w.r.t. the applied
  lenses
\item
  What should or could be?
\item
  How to get from what was to what should be?
\end{enumerate}

A comparison of the meaning (what was) and the goals (what should be) is
the basis for this. The final question can be used to find ideas on how
the involvement of the external stakeholders can be improved. These
proactive measures can be applied in all facets of the preparation and
the action.

\end{document}